\documentclass[11pt,aps,prd,nofootinbib,superscriptaddress,preprint,preprintnumbers
]{revtex4-2}

\usepackage{amssymb}
\usepackage{amsmath}
\usepackage{bbm}
\usepackage{graphicx}
\usepackage{hyperref}
\usepackage{multirow}
\usepackage{booktabs}
\usepackage{array}

\allowdisplaybreaks[2]

\DeclareMathOperator{\dlog}{\mathit{d}log}
\newcommand{\sss}{\scriptscriptstyle}
\newcommand{\nn}{\nonumber \\}
\newcommand{\dA}{d\mathbb{A}}
\newcommand{\KK}{\mathbb{K}}
\newcommand{\vpq}{\vphantom{m_q^2}}
\newcommand{\SKLP}{State Key Laboratory of Particle Detection and Electronics, University of Science and Technology of China, Hefei 230026, Anhui, People’s Republic of China}
\newcommand{\USTC}{Department of Modern Physics, University of Science and Technology of China, Hefei 230026, Anhui, People's Republic of China}
\newcommand{\ACFS}{Anhui Center for Fundamental Sciences in Theoretical Physics, University of Science and Technology of China, Hefei 230026, Anhui, People’s Republic of China}
\newcommand{\NUS}{Department of Physics, National University of Singapore, Singapore 117551, Singapore}

\begin{document}

\title{\boldmath Next-to-next-to-leading order $\text{QCD} \otimes \text{EW}$ corrections to $Z$-boson pair production at electron-positron colliders}

\author{Zhe Li}
\affiliation{\SKLP}
\affiliation{\USTC}

\author{Ren-You Zhang}
\email{zhangry@ustc.edu.cn}
\affiliation{\SKLP}
\affiliation{\USTC}
\affiliation{\ACFS}

\author{Shu-Xiang Li}
\affiliation{\SKLP}
\affiliation{\USTC}

\author{Xiao-Feng Wang}
\affiliation{\SKLP}
\affiliation{\USTC}

\author{Pan-Feng Li}
\affiliation{\SKLP}
\affiliation{\USTC}

\author{Yi Jiang}
\affiliation{\SKLP}
\affiliation{\USTC}

\author{Liang Han}
\affiliation{\SKLP}
\affiliation{\USTC}

\author{Qing-hai Wang}
\affiliation{\NUS}

\date{\today}

\begin{abstract}
We present a comprehensive analytic calculation of the next-to-next-to-leading order $\text{QCD} \otimes \text{EW}$ corrections to $Z$-boson pair production at electron-positron colliders. The two-loop master integrals essential to this calculation are evaluated using the differential equation method. In this work, we detail the formulation and solution of the canonical differential equations for the two-loop three-point master integrals with two on-shell $Z$-boson external legs and a massive internal quark in the loops. These canonical master integrals are systematically expanded as Taylor series in the dimensional regulator, $\epsilon = (4-d)/2$, up to the order of $\epsilon^4$, with coefficients expressed in terms of Goncharov polylogarithms up to weight four. Upon applying our analytic expressions of these master integrals to the phenomenological analysis of $Z$-pair production, we observe that the $\mathcal{O}(\alpha \alpha_s)$ corrections manifest at a level of approximately one percent compared to the leading-order predictions, underscoring their significance for comparisons with future high-precision experimental data.

\begin{description}
\item[keywords]
$Z$-boson pair production, NNLO $\text{QCD} \otimes \text{EW}$ corrections, Canonical differential equations, Goncharov polylogarithms
\end{description}
\end{abstract}

\maketitle

\section{\label{sec:1}Introduction}

\par
The discovery of the Higgs boson \cite{ATLAS:2012yve,CMS:2012qbp} at the CERN Large Hadron Collider (LHC) represents a landmark validation of the Standard Model (SM). Nevertheless, achieving a thorough and precise understanding of the SM is essential for unraveling the mechanism of electroweak (EW) symmetry breaking and advancing our search for new physics beyond the SM. In this pursuit, the study of neutral gauge boson pair production assumes a pivotal role. It not only sheds light on the study of the EW gauge structure and the exploration of triple gauge couplings (TGCs), but also serves as a crucial avenue for uncovering traces of new physics. Additionally, $Z$-boson pair production is an irreducible background to the production of a Higgs boson decaying into two neutral vector bosons. The production of $Z$-boson pair has been comprehensively studied in $e^+e^-$ collisions at LEP \cite{ALEPH:1999ytd,L3:2003kow,DELPHI:2003kgg,ALEPH:2006bhb,DELPHI:2007gzg,ALEPH:2009rug}, in $p\bar{p}$ collisions at Tevatron \cite{CDF:2008eiu,D0:2008sin,CDF:2014nef}, and in $pp$ collisions at LHC \cite{CMS:2021pqj,ATLAS:2011ahl,CMS:2013piy,CMS:2015qgb,ATLAS:2016bxw,ATLAS:2015guq,CMS:2016ogx,CMS:2020gtj}. Proposals for several future lepton colliders, including the International Linear Collider (ILC) \cite{Behnke:2013xla,ILC:2013jhg,Bambade:2019fyw}, the Circular Electron-Positron Collider (CEPC) \cite{CEPCStudyGroup:2018rmc,CEPCStudyGroup:2018ghi} and the Future Circular Collider (FCC-ee) \cite{FCC:2018byv,FCC:2018evy}, hold immense for advancing our comprehension of the EW interactions within the SM and probing potential candidates for new physics beyond the SM. These lepton colliders, benefiting from cleaner experimental environments compared to hadron colliders, are anticipated to achieve measurements with permille precision. Therefore, it is imperative that the theoretical predictions match or exceed the same level of precision. Extensive research has been devoted to studying the leading order (LO) predictions and the next-to-leading order (NLO) corrections for the $e^+e^- \rightarrow ZZ$ process over the past few decades \cite{Brown:1978mq,Gaemers:1978hg,Denner:1988tv,Denner:1988nq,Gounaris:2002fa,Gounaris:2002za,Demirci:2022lmr,Bondarenko:2024txj}. However, the next-to-next-to-leading order (NNLO) corrections to $e^+e^- \rightarrow ZZ$ still remain elusive. These NNLO corrections consist of two components: the pure EW $\mathcal{O}(\alpha^2)$ corrections and the mixed $\text{QCD} \otimes \text{EW}$ $\mathcal{O}(\alpha \alpha_s)$ corrections. The calculation of the pure EW corrections poses a significant challenge due to the tremendous number of two-loop Feynman diagrams with multiple scales. In contrast, the mixed $\text{QCD} \otimes \text{EW}$ corrections, though intricate, are more tractable, and may hold greater numerical significance due to the relatively large QCD coupling constant. In light of these considerations, this paper focuses on the analytic calculation of the NNLO $\text{QCD} \otimes \text{EW}$ corrections to $e^+e^- \rightarrow ZZ$.

\par
All mixed $\text{QCD} \otimes \text{EW}$ two-loop Feynman diagrams for $e^+e^- \rightarrow ZZ$ are factorizable and originate from the corrections to TGCs. Upon implementing the tensor reduction procedure, the two-loop scattering amplitude is expressed as a linear combination of scalar Feynman integrals, which can be categorized into several families. Within each family, the scalar integrals are interrelated and can be systematically reduced to a set of master integrals (MIs) using integration-by-parts (IBP) identities \cite{Tkachov:1981wb,Chetyrkin:1981qh}. We conduct analytic calculations of these two-loop MIs utilizing the canonical differential equation method \cite{Henn:2013pwa,Henn:2014qga}. Leveraging \emph{leading singularities} \cite{Henn:2013pwa,Henn:2014qga,Dlapa:2021qsl,He:2022ctv,Weinzierl:2022eaz}, \emph{dlog integrals} \cite{Chicherin:2018old,Henn:2020lye}, \emph{Magnus exponential} \cite{magnus1954exponential,blanes2009magnus,Argeri:2014qva,Weinzierl:2022eaz}, and various other techniques, we can establish a proper basis comprising a set of uniform transcendental (UT) integrals that satisfy the canonical differential equations. Furthermore, we present \emph{Chen's iterated integral} \cite{chen1977iterated} solution of the canonical differential system in terms of Goncharov polylogarithms (GPLs) \cite{Goncharov:1998kja}, following the rationalization of all square roots appearing in the system. Ultimately, we apply our analytic expressions of the canonical MIs to the computation of the NNLO $\text{QCD} \otimes \text{EW}$ corrections to $e^+e^- \rightarrow ZZ$, and provide detailed numerical predictions for both integrated and differential cross sections.

\par
Before delving into further analyses, it is essential to provide an overview of the current state-of-the-art in analytic studies of two-loop Feynman integrals, stemming from the NNLO $\text{QCD} \otimes \text{EW}$ corrections. The two-loop two-point MIs involved in the $\text{QCD} \otimes \text{EW}$ two-loop vector-boson self-energies have been thoroughly studied in Refs.\cite{Chang:1981qq,Djouadi:1987gn,Djouadi:1987di,Kniehl:1988ie,Kniehl:1989yc,Djouadi:1993ss}. The analytic results of the two-loop triangle MIs for the QCD corrections to $H \rightarrow V^{\ast}V^{\ast}$ ($V=W, Z$)  with massless internal particles and arbitrary external momenta have been effectively expressed in terms of GPLs in Refs.\cite{Usyukina:1994iw,Birthwright:2004kk,Chavez:2012kn}. These results can be employed in the calculation of the massless two-loop three-point MIs involved in the $\mathcal{O}(\alpha\alpha_s)$ corrections to the $V^{\ast}ZZ$ vertex. In addition, all two-loop three-point MIs necessary for the QCD corrections to $H \rightarrow \gamma \gamma$ and $H \rightarrow Z \gamma$ with a massive top-quark loop have been expressed in terms of harmonic polylogarithms \cite{Fleischer:2004vb,Harlander:2005rq,Aglietti:2006tp} and GPLs \cite{Bonciani:2015eua,Gehrmann:2015dua}, respectively. Moreover, all two-loop triangle MIs for QCD corrections to the interaction vertex of an off-shell neutral boson with a pair of $W$ bosons are well documented in terms of GPLs \cite{DiVita:2017xlr,Ma:2021cxg,Li:2024dlh}. A canonical set of two-loop triangle MIs, contributing to the $\mathcal{O}(\alpha\alpha_s)$ corrections to the $HZV^{\ast}$ ($V=Z, \gamma$) vertex and $H \rightarrow ZZ^{\ast}$ decay, has been studied in Refs.\cite{Wang:2019fxh,Chaubey:2022hlr}. The MIs for $H \rightarrow ZZ^{\ast}$, characterized by a massive internal loop and three distinct external momentum squares, are expressed in terms of Chen's iterated integrals with dlog one-forms that contain a residual non-rationalizable square root \cite{Chaubey:2022hlr}. The two-loop triangle MIs with two on-shell $Z$-boson legs and a massive internal loop, arising from the NNLO $\text{QCD} \otimes \text{EW}$ corrections to $e^+e^- \rightarrow ZZ$, can be considered a special case of the two-loop MIs in the $H \rightarrow ZZ^{\ast}$ decay, where the two $Z$-boson external legs are forced on their mass-shell. We analytically calculate these two-loop three-point MIs associated with two on-shell $Z$-boson legs, and present them in a more user-friendly format, namely GPLs. This representation offers a faster, more accurate, and more convenient numerical evaluation compared to the iterated integrals with dlog one-forms.

\par
The rest of this paper is structured as follows. In Sec.\ref{sec:2}, we introduce our notations and conventions, present the framework for our calculation, delineate the workflow of our computation, and engage a discussion concerning the $\gamma^5$ schemes. In Sec.\ref{sec:3}, we dedicate to the elaboration on the analytic calculation for the two-loop three-point MIs originating from the $\mathcal{O}(\alpha \alpha_s)$ corrections to $e^+e^- \rightarrow ZZ$. In Sec.\ref{sec:4}, we utilize the analytic MIs derived in Sec.\ref{sec:3} to compute the NNLO $\text{QCD} \otimes \text{EW}$ corrected integrated and differential cross sections for $Z$-boson pair production at electron-positron colliders. Finally, we conclude with a concise summary in Sec.\ref{sec:5}.

\section{\label{sec:2}Calculation strategy}

\subsection{\label{sec:2A}General setup}

\par
In this paper, we study the NNLO $\text{QCD} \otimes \text{EW}$ corrections to the scattering process
\begin{equation}
e^+(p_1) + e^-(p_2) \rightarrow Z(p_3) + Z(p_4)\,.
\end{equation}
All the external particles are on their mass-shell, i.e., $p_1^2 = p_2^2 = m_e^2$ and $p_3^2 = p_4^2 = m_{\sss Z}^2$. The Mandelstam invariants for this process are defined by
\begin{equation}
s = (p_1 + p_2)^2\,,
\qquad
t = (p_1 - p_3)^2\,,
\qquad
u = (p_1 - p_4)^2\,,
\end{equation}
which satisfy $s + t + u = 2\, (m_e^2 + m_{\sss Z}^2)$ due to energy-momentum conservation. The four-momentum of the final-state $Z$ boson can be parameterized as
\begin{equation}
p_{3}
=
(E_3,\, \vec{p}_3)
=
\frac{\sqrt{s}}{2}\,
(1,\, \beta \sin\theta \cos\phi,\, \beta \sin\theta \sin\phi,\, \beta\cos\theta)\,,
\end{equation}
where $\beta$ is the magnitude of momentum normalized by $\sqrt{s}/2$, $\theta$ the polar angle with respect to the direction of the incident positron beam, and $\phi$ the azimuthal angle around the positron beam axis. In the center-of-mass frame, $\beta$ is given by
\begin{equation}
\beta
=
\sqrt{ 1 - \frac{4\, m_{\sss Z}^2}{s}}\,.
\end{equation}
Leptons and light quarks $(u, d, c, s)$ are treated as massless particles throughout our calculation unless explicitly stated otherwise. Then, the Mandelstam variables $t$ and $u$ can be expressed as
\begin{equation}
t,u = m_{\sss Z}^2 - \frac{s}{2}\,(1 \mp \beta\cos\theta)\,.
\end{equation}
The lowest-order unpolarized differential cross section for this $2 \rightarrow 2$ scattering process is simply given by
\begin{equation}
\mathrm{d} \sigma_{\sss{\mathrm{LO}}}
=
\frac{\beta}{128\, \pi^{2} s}\, \frac{1}{4}\,
\sum\limits_{\text{spin}}\, \left| \mathcal{M}_{0}(s, t) \right|^{2}
\mathrm{d}\cos\theta \,\mathrm{d}\phi\,,
\end{equation}
where $\mathcal{M}_{0}$ is the tree-level Feynman amplitude. Throughout our calculation, we adopt the 't Hooft-Feynman gauge.

\par
The on-shell (OS) renormalization scheme \cite{Denner:1991kt,Denner:2019vbn} is employed to remove the ultraviolet (UV) divergences, which are regularized by using dimensional regularization (DR) \cite{tHooft:1972tcz,Bollini:1972ui} in both one-loop and two-loop calculations. An infinitesimal fictitious photon mass is introduced to regularize the infrared (IR) divergences induced by the virtual photon in loops, which can be canceled exactly by those from the real photon emission. We adopt the phase space slicing method \cite{Dittmaier:1999mb,Denner:2000bj,Harris:2001sx} to extract the IR singularities from the real photon emission. Additionally, the initial-state QED corrections are implemented in the leading-logarithmic approximation with the structure-function method \cite{Beenakker:1996kt,Denner:2000bj}. The $G_{\mu}$ scheme is adopted for the renormalization of electric charge, wherein the fine structure constant can be derived from the Fermi constant $G_{\mu}$ via
\begin{equation}
\alpha_{\sss G_\mu}
=
\frac{\sqrt{2}\,G_\mu m_{\sss W}^2}{\pi}
\Big(1 - \frac{m_{\sss W}^2}{m_{\sss Z}^2} \Big)\,.
\end{equation}
It is important to note that the corresponding charge renormalization constant in the $G_{\mu}$ scheme should be modified to $\delta Z_{e,\sss{G_\mu}} = \delta Z_{e,{\alpha(0)}} - \Delta r/2$, where the subtraction term $\Delta r$ comprises the higher-order corrections to muon decay \cite{Sirlin:1980nh,Denner:2019vbn}. The explicit expression for the renormalization constant in the $\alpha(0)$ scheme, $\delta Z_{e,{\alpha(0)}}$, has been detailed discussed in Refs.\cite{Denner:1991kt,Denner:2019vbn}. For the computation of the LO cross section and NLO EW corrections, we employ the modified \texttt{FormCalc} and \texttt{LoopTools} packages \cite{Hahn:1998yk,vanOldenborgh:1990yc}. To validate the accuracy and reliability of our NLO results, we use our developed computational tools to conduct rigorous numerical cross-checks against the benchmarks provided in Ref.\cite{Demirci:2022lmr}, confirming excellent concordance. For a comprehensive elucidation of the NLO EW corrections to $e^+e^- \rightarrow ZZ$, please refer to Refs.\cite{Gounaris:2002fa,Demirci:2022lmr,Bondarenko:2024txj}.

\par
It is well known that the cyclicity of Dirac trace conflicts with the anticommutativity relation $\{\gamma^5,\, \gamma^{\mu}\} = 0$ in the framework of DR. Consequently, there are two prominent strategies for the practical treatment of $\gamma^5$ in DR: either preserving or disregarding the anticommutativity relation. To circumvent the challenging nontrivial renormalizations and the computational complexities associated with the 't Hooft-Veltman scheme \cite{tHooft:1972tcz} and its subsequent iterations \cite{Breitenlohner:1977hr,Aoyama:1980yw,Bonneau:1990xu,Barroso:1990ti,Larin:1993tq}, wherein the anticommutativity relation is sacrificed, we adopt the K{\"o}rner-Kreimer-Schilcher (KKS) scheme \cite{Kreimer:1989ke,Korner:1991sx,Kreimer:1993bh} and its reformulated version \cite{Chen:2022vzo,Chen:2023lus}. Within the KKS scheme, the cyclicity of the Dirac trace with odd numbers of $\gamma^5$ is deliberately abandoned, prioritizing the preservation of the anticommutativity property of $\gamma^5$. To address the ambiguity arising from the dependence of the Dirac trace on the reading start point in a fermion chain, a specific reading prescription is introduced. In this prescription, the final expression of the non-cyclic $\gamma^5$ trace is defined as the average of all possible reading points, starting from both the head and the tail of an axial-vector three-point fermion-gauge vertex that contains the maximal one-particle-irreducible non-singlet-type loop correction \cite{Kreimer:1993bh,Chen:2023lus}. Detailed technical descriptions for the KKS scheme and its reformulated version can be found in Refs.\cite{Kreimer:1993bh,Chen:2023lus}. For additional discussions on $\gamma^5$ schemes in $d = 4- 2\, \epsilon$ dimensions, we refer readers to Ref.\cite{Jegerlehner:2000dz}, which provides deeper insights and comprehensive considerations.

\subsection{\label{sec:2B}NNLO $\text{QCD} \otimes \text{EW}$ corrections}

\par
\begin{figure}[htbp]
\centering
\includegraphics[width = 0.95\textwidth]{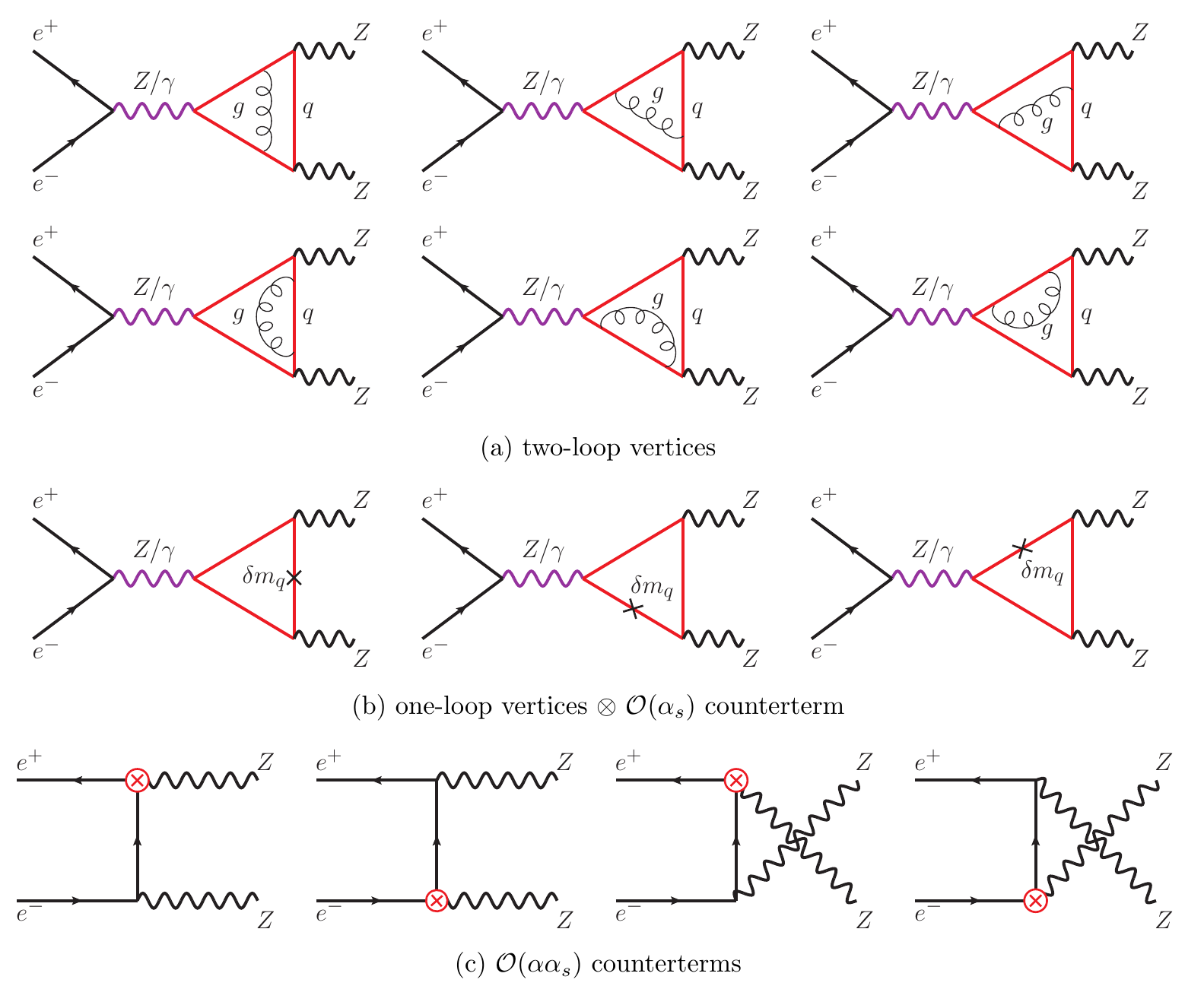}
\caption{Representative Feynman diagrams for NNLO $\text{QCD} \otimes \text{EW}$ corrections to $e^+ e^- \rightarrow ZZ$. The black crosses symbolize the heavy quark mass counterterm at $\mathcal{O}(\alpha_s)$, and the red circled crosses signify the vertex counterterm at $\mathcal{O}(\alpha\alpha_s)$.}
\label{fig1}
\end{figure}

The NNLO $\text{QCD} \otimes \text{EW}$ $\mathcal{O}(\alpha \alpha_s)$ corrections to $e^+ e^- \rightarrow ZZ$ are categorized into the following three distinct contributions:
\begin{enumerate}
\item Two-loop vertex corrections induced by attaching one gluon to each one-loop quark line in every possible way.
\item One-loop vertex corrections with insertions of $\mathcal{O}(\alpha_s)$ quark mass counterterm\footnote{Due to the QED-like Ward identity, the vertex and fermion wave-function counterterms precisely cancel each other, leaving contributions solely from quark mass renormalization.}.
\item $\mathcal{O}(\alpha\alpha_s)$ vertex counterterms.
\end{enumerate}
Some representative Feynman diagrams for these $\mathcal{O}(\alpha\alpha_s)$ corrections are depicted in Fig.\ref{fig1}. The $\mathcal{O}(\alpha_s)$ quark mass renormalization constant $\delta m_q$ in the OS scheme is given by \cite{Bernreuther:2004ih}
\begin{equation}
\delta m_q
=
-\, m_q\, \frac{\alpha_s}{2\pi}\,
C(\epsilon)\,
\Big(\frac{\mu^2}{m_{q}^2}\Big)^\epsilon\,
\frac{C_F}{2}\,
\frac{3-2\,\epsilon}{\epsilon\,(1-2\,\epsilon)}\,,
\end{equation}
where $C(\epsilon) = (4\pi)^{\epsilon}\, \Gamma(1+\epsilon)$ and $C_F = 4/3$. The explicit expressions for the $\mathcal{O}(\alpha \alpha_s)$ renormalization constants can be derived by substituting the one-loop gauge-boson self-energies in the NLO EW renormalization constants \cite{Denner:1991kt} with their corresponding two-loop $\mathcal{O}(\alpha\alpha_s)$ counterparts \cite{Dittmaier:2015rxo}.

\par
We employ a comprehensive set of tools and methodologies to calculate the NNLO $\text{QCD} \otimes \text{EW}$ corrections to $e^+e^- \rightarrow ZZ$. Initially, \texttt{FeynArts} \cite{Hahn:2000kx} is used to generate the Feynman diagrams and amplitudes for the $Z$-pair production process. Then, we utilize an in-house \texttt{Mathematica} code based on \texttt{FeynCalc} \cite{Mertig:1990an,Shtabovenko:2020gxv} to manipulate the Dirac algebra and express the two-loop amplitudes as linear combinations of numerous scalar Feynman integrals, which are not independent. To reduce these extensive Feynman integrals to a small set of irreducible MIs, we use the publicly available package \texttt{Kira} \cite{Maierhofer:2017gsa,Klappert:2020nbg}, which implements the Laporta algorithm \cite{Laporta:2000dsw} for solving IBP identities. Subsequently, we analytically calculate all MIs appearing in the NNLO $\text{QCD} \otimes \text{EW}$ corrections using the canonical differential equation method and express them in terms of GPLs. Additionally, several \texttt{Mathematica} ancillary files containing the analytic expressions of the MIs are provided as supplementary materials accompanying the arXiv submission of this paper. To validate our analytical results, numerical checks for the MIs are conducted using the \texttt{AMFlow} \cite{Liu:2017jxz,Liu:2022chg} package, showing perfect agreement between our analytical and numerical results. Detailed elaboration on the analytic calculation of MIs is presented in Sec.\ref{sec:3}.

\section{\label{sec:3}Canonical differential equations}

\par
In this section, we concentrate on the analytic calculation of the massive two-loop triangle MIs for the NNLO $\text{QCD} \otimes \text{EW}$ corrections to the $e^+ e^- \rightarrow ZZ$ process.

\subsection{\label{sec:3A}Integral family}

\par
\begin{figure}[htbp]
\centering
\includegraphics[width = 1.0\textwidth]{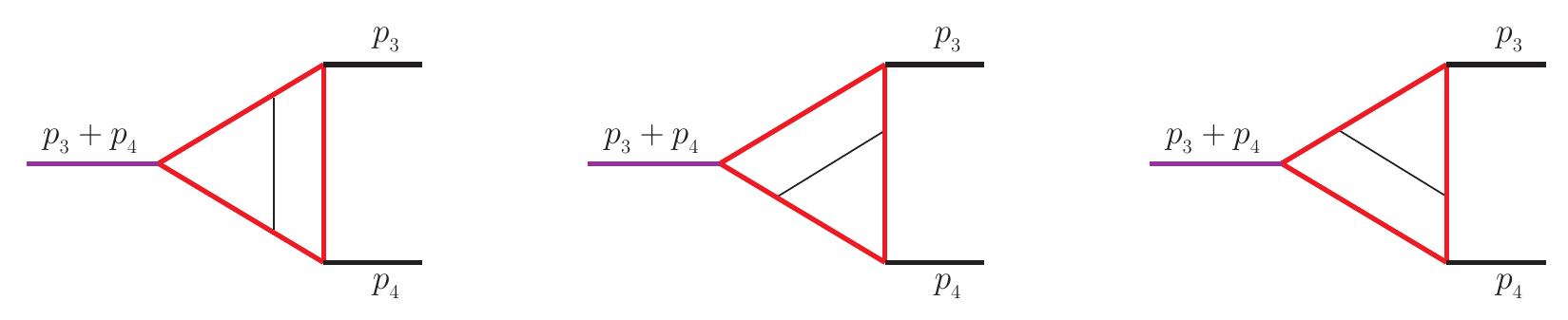}
\caption{Three top-level topologies of the integral family $\mathcal{F}$. The red lines represent massive propagators, while the thin black lines denote massless propagators. The purple lines indicate an external off-shell leg with momentum squared $s$, whereas the black external lines signify the on-shell $Z$ bosons.}
\label{fig2}
\end{figure}

The integral family $\mathcal{F}$ of the two-loop three-point MIs with two on-shell $Z$ boson legs and a massive internal quark loop, as depicted in the panel (a) of Fig.\ref{fig1}, is defined by
\begin{equation}
I(\alpha_1, \ldots, \alpha_7)
=
\int \mathcal{D}^d l_1 \mathcal{D}^d l_2\,
\frac{1}{D_1^{\alpha_1} \ldots D_7^{\alpha_7}}\,,
\end{equation}
where $d=4-2\,\epsilon$ is the spacetime dimension, and the integration measure is defined as
\begin{equation}
\label{eq:measure}
\mathcal{D}^d l_i
=
\frac{d^dl_i}{(2\pi)^d}\, \Big( \frac{i S_{\epsilon}}{16\pi^2} \Big)^{-1} \big(m_q^2\big)^{\epsilon}
\qquad
\text{with}
\qquad
S_{\epsilon} = (4\pi)^{\epsilon}\, \Gamma(1+\epsilon)\,.
\end{equation}
The integral family $\mathcal{F}$ is identified by the following set of propagators,
\begin{align}
&
D_1 = l_1^2 - m_q^2\,,
&\quad&
D_3 = (l_1+p_3)^2-m_q^2\,,
&\quad&
D_5 = (l_1-p_4)^2 - m_q^2\,,
&\quad&
D_7 = (l_1-l_2)^2
\nonumber \\
&
D_2 = l_2^2 - m_q^2\,,
&\quad&
D_4 = (l_2+p_3)^2-m_q^2\,,
&\quad&
D_6 = (l_2-p_4)^2 - m_q^2\,,
\end{align}
with $p_3^2=p_4^2=m_{\sss Z}^2$. The three topologies illustrated in Fig.\ref{fig2} correspond to the following three top-sectors of the integral family $\mathcal{F}$,
\begin{equation}
[1, 0, 1, 1, 1, 1, 1]\,,
\qquad\quad
[1, 1, 0, 1, 1, 1, 1]\,,
\qquad\quad
[1, 1, 1, 1, 0, 1, 1]\,.
\end{equation}
In this section, we elaborate on the construction of the canonical basis for the integral set $\mathcal{S}$ induced by the three top-sectors,
\begin{equation}
\mathcal{S}
=
[1, 0, 1, 1, 1, 1, 1]_{\digamma}
\cup\,
[1, 1, 0, 1, 1, 1, 1]_{\digamma}
\cup\,
[1, 1, 1, 1, 0, 1, 1]_{\digamma}\,,
\end{equation}
where the subscript $\digamma$ is defined by
\begin{equation}
[s_1, \ldots, s_n]_{\digamma}
=
\bigcup_{s_i^{\prime} \leqslant s_i}
[s_1^{\prime}, \ldots, s_n^{\prime}]\,.
\end{equation}
For brevity and clarity, the three top-sectors are denoted as $\mathcal{I}_{134567}$, $\mathcal{I}_{124567}$ and $\mathcal{I}_{123467}$, with the sub-sectors of these top-sectors adhering to the same naming convention. Following the IBP reduction, we derive a set of $26$ MIs for the integral set $\mathcal{S}$. In the subsequent Sec.\ref{sec:3B}, we focus on the analytic calculation of these MIs using the method of canonical differential equations.

\subsection{\label{sec:3B}Canonical differential equations for MIs}

\par
To begin with, we establish a system of standard differential equations,
\begin{equation}
d{\mathbf{I}}(\vec{x},\epsilon)
=
d\tilde{\mathbb{A}}(\vec{x},\epsilon)\, {\mathbf{I}}(\vec{x},\epsilon)\,,
\end{equation}
for a pre-canonical basis ${\mathbf{I}}(\vec{x},\epsilon)$ that comprises $26$ MIs derived using the Laporta algorithm. This differential system is generated by taking partial derivatives of the MIs with respect to the kinematic variables ($s, m_{\sss Z}^2, m_t^2$) and employing IBP identities for scalar reduction, facilitated by \texttt{LiteRed} \cite{Lee:2012cn,Lee:2013mka} and \texttt{Kira}. As suggested in Refs.\cite{Henn:2013pwa,Henn:2014qga}, we can transform the pre-canonical basis ${\mathbf{I}}(\vec{x},\epsilon)$ into a canonical basis ${\mathbf{g}}(\vec{x},\epsilon)$, satisfying the following canonical differential equations,
\begin{equation}
\label{eq:canonicalDE}
d{\mathbf{g}}(\vec{x},\epsilon)
=
\epsilon\, d{\mathbb{A}}(\vec{x})\, \mathbf{g}(\vec{x},\epsilon)
=
\epsilon\, \Big[ \sum_{i=1}^{n} \mathbb{M}_i \dlog \eta_{i}(\vec{x}) \Big]\, \mathbf{g}(\vec{x},\epsilon)\,,
\end{equation}
where $\mathbb{M}_i$ are constant matrices and the symbol letters $\eta_i$ are algebraic functions of $\vec{x}$. The general solution to the canonical differential equations \eqref{eq:canonicalDE} can be expressed in terms of Chen's iterated integrals \cite{chen1977iterated}, formulated as the following path-ordered exponential,
\begin{equation}
\mathbf{g}(\vec{x},\epsilon)
=
\mathcal{P} \exp \Big(\epsilon \int_\gamma \dA\Big)\,
\mathbf{g}(\vec{x}_0, \epsilon)\,,
\end{equation}
where $\mathcal{P}$ stands for the path ordering along the integration path $\gamma$ from the start point $\vec{x}_0$ to the point of interest $\vec{x}$, and $\mathbf{g}(\vec{x}_0,\epsilon)$ is the vector of boundary constants at $\vec{x}_0$. The path-ordered exponential is a shorthand notation for the following series:
\begin{equation}
\label{eq:pathorderedexp}
\mathcal{P} \exp\Big(\epsilon \int_\gamma \dA\Big)
=
\mathbbm{1}  + \epsilon \int_\gamma \dA
+ \epsilon^2 \int_{\gamma} \dA\, \dA
+ \epsilon^3 \int_\gamma \dA\, \dA\, \dA + \cdots 
\end{equation}
The $n$-th term in this expansion represents an $n$-fold iterated integral,
\begin{equation}
\int_{\gamma}\underbrace{\dA \cdots \dA}_{n\,\text{times}}
=
\int_0^1 dt_1\, \KK(t_1) \int_0^{t_1} dt_2 \, \KK(t_2)\,  \cdots
\int_0^{t_{n-1}} dt_n \,\KK(t_n)\,,
\end{equation}
where $\mathbb{K}(t)\, dt$ is the pullback of the matrix-valued differential one-form $d \mathbb{A}$ to the unit interval $[0, 1]$. It is crucial to note that these iterated integrals do not depend on the choice of the integration path $\gamma$, unless it encounters the singularities or branch cuts of $d\mathbb{A}$.

\par
We adopt a ``bottom-up" sector-wise methodology to construct the canonical differential equations by arranging all MIs sequentially from lower to higher sectors and then systematically transforming them into the canonical basis, sector by sector. We employ Baikov leading singularity analysis techniques \cite{Dlapa:2021qsl,He:2022ctv,Weinzierl:2022eaz} to construct the UT integrals or candidate UT integrals for a given sector, aided by the publicly available package \texttt{DlogBasis} \cite{Henn:2020lye}. Furthermore, if required, the contributions arising from sub-sectors are accounted for to transform the off-diagonal blocks into canonical forms. By combining with the application of the \emph{Magnus exponential} method \cite{magnus1954exponential,blanes2009magnus,Argeri:2014qva}, we derive the following canonical basis,
{
\fontsize{11pt}{16}\selectfont
\begin{align}
\label{eq:UTbasis}
&\mathcal{I}_{36} :
&g_1    = {}
                      &\epsilon^2\, I_{0,0,2,0,0,2,0} \,, &
&\mathcal{I}_{346} :
&g_2    = {}
                      &\lambda_1\, \epsilon^2\, I_{0,0,2,1,0,2,0} \,,
\nn
&\mathcal{I}_{136} :
&g_3    = {}
                      &\lambda_2\, \epsilon^2\, I_{1,0,2,0,0,2,0} \,, &
&\mathcal{I}_{3456} :
&g_4    = {}
                      &\lambda_1^2\, \epsilon^2\, I_{0,0,2,1,1,2,0} \,,
\nn
&\mathcal{I}_{1346} :
&g_5    = {}
                      &\lambda_1\, \lambda_2\, \epsilon^2\, I_{1,0,2,1,0,2,0} \,, &
&\mathcal{I}_{1245} :
&g_6    = {}
                      &\lambda_2^2\, \epsilon^2\, I_{1,1,0,2,2,0,0} \,,
\nn
&\mathcal{I}_{1345} :
&g_7    = {}
                      &\lambda_3\, \epsilon^3\, I_{1,0,1,2,1,0,0} \,, &
&\mathcal{I}_{12456} :
&g_8    = {}
                      &\lambda_2\, \lambda_3\, \epsilon^3\, I_{1,1,0,1,2,1,0} \,, 
\nn
&\mathcal{I}_{13456} :
&g_9    = {}
                      &\lambda_1\, \lambda_3\, \epsilon^3\, I_{1,0,1,1,1,2,0} \,,
\nn
&\mathcal{I}_{367} :
&g_{10} = {}
                      &s\, \epsilon^2\, I_{0,0,2,0,0,2,1} \,,
\nn
&
&g_{11} = {}
                      &\lambda_1\, \big(
                       \epsilon^2\, I_{0,0,2,0,0,1,2}
                       + \frac{1}{2}\, \epsilon^2\, I_{0,0,2,0,0,2,1}
                       \big) \,,
\nn
&\mathcal{I}_{167} :
&g_{12} = {}
                      &m_{\sss Z}^2\, \epsilon^2\, I_{2,0,0,0,0,2,1} \,,
\nn
&
&g_{13} = {}
                      &\lambda_2\, \big(
                       \epsilon^2\, I_{2,0,0,0,0,1,2}
                       + \frac{1}{2}\, \epsilon^2\, I_{2,0,0,0,0,2,1}
                       \big) \,,
\nn
&\mathcal{I}_{1467} :
&g_{14} = {}
                      &\lambda_3\, \epsilon^3\, I_{2,0,0,1,0,1,1} \,,
\nn
&
&g_{15} = {}
                      &\lambda_3\, m_q^2\, \epsilon^2\, I_{3,0,0,1,0,1,1} \,,
\nn
&
&g_{16} = {}
                      &\rlap{$\displaystyle
                       \lambda_1\, \big[\,
                       ( m_q^2 - m_{\sss Z}^2)\, \epsilon^2\, I_{2,0,0,1,0,2,1}
                       + m_q^2\, \epsilon^2\, I_{3,0,0,1,0,1,1}
                       + \frac{3}{2}\, \epsilon^3\, I_{2,0,0,1,0,1,1}
                       \,\big] \,,
                             $}
\nn
&\mathcal{I}_{1367} :
&g_{17} = {}
                      &\lambda_3\, \epsilon^3\, I_{1,0,1,0,0,2,1} \,,
\nn
&
&g_{18} = {}
                      &\lambda_3\, m_q^2\, \epsilon^2\, I_{1,0,1,0,0,3,1} \,,
\nn
&
&g_{19} = {}
                      &\rlap{$\displaystyle
                       \frac{\lambda_2 }{4\, m_{\sss Z}^4 + 8\, s\, m_q^2 - 12\, m_q^2\, m_{\sss Z}^2}
                       \Big\{\,
                       4\, \big[\, s\, ( s\,  m_q^2 - 3\, m_q^2\, m_{\sss Z}^2 + m_{\sss Z}^4 ) + m_q^4\, m_{\sss Z}^2 \,\big]\, \epsilon^2\, I_{1,0,2,0,0,2,1}
                             $}
\nn
&
&
                      &-\rlap{$\displaystyle
                       4\, m_q^2\, \big[\, s\, ( m_q^2 + m_{\sss Z}^2 ) - 2\, m_q^2\, m_{\sss Z}^2 \,\big]\, \epsilon^2\, I_{1,0,1,0,0,3,1}
                             $}
\nn
&
&
                      &-6\, \big[\, s\, ( m_q^2 + m_{\sss Z}^2 ) - 2\, m_q^2\, m_{\sss Z}^2 \,\big]\, \epsilon^3\, I_{1,0,1,0,0,2,1}
\nn
&
&
                      &-\rlap{$\displaystyle
                       3\, m_{\sss Z}^2\, ( m_q^2  - m_{\sss Z}^2 )\, \epsilon^2\, I_{2,0,0,0,0,2,1}
                       + 3\, s\, ( m_q^2 - m_{\sss Z}^2 )\, \epsilon^2\, I_{0,0,2,0,0,2,1}
                       \Big\} \,,
                             $}
\nn
&\mathcal{I}_{13467} :
&g_{20} = {}
                      &\lambda_3\, \epsilon^4\, I_{1,0,1,1,0,1,1} \,,
\nn
&
&g_{21} = {}
                      &\lambda_1\, \lambda_3\, \epsilon^3\, I_{1,0,1,1,0,2,1} \,,
\nn
&
&g_{22} = {}
                      &\lambda_2\, \lambda_3\, \epsilon^3\, I_{2,0,1,1,0,1,1} \,,
\nn
&
&g_{23} = {}
                      &\rlap{$\displaystyle
                       s\, \big[\, m_q^2\, ( s - 4\, m_{\sss Z}^2 ) + m_{\sss Z}^4 \,\big]\, \epsilon^2\, I_{2,0,1,1,0,2,1}
                       - s\, m_{\sss Z}^2\, \epsilon^3\, I_{2,0,1,1,0,1,1}
                             $}
\nn
&
&
                      &- \rlap{$\displaystyle
                       2\, s\, ( m_{\sss Z}^2 - 2\, m_q^2 )\, \epsilon^2\, I_{1,0,2,1,0,2,0}
                       + s\, ( s - 2\, m_{\sss Z}^2 )\, \epsilon^3\, I_{1,0,1,1,0,2,1} \,,
                               $}
\nn
&\mathcal{I}_{12457} :
&g_{24} = {}
                      &\lambda_3\, \epsilon^4\, I_{1,1,0,1,1,0,1} \,,
\nn
&
&g_{25} = {}
                      &\lambda_2\, \lambda_3\, \epsilon^3\, I_{1,1,0,1,2,0,1} \,,
\nn
&
&g_{26} = {}
                      &\rlap{$\displaystyle
                       s\, \big[\, m_q^2\, ( s - 4\, m_{\sss Z}^2 ) + m_{\sss Z}^4 \,\big]\, \epsilon^2\, I_{1,1,0,2,2,0,1}
                       - 2\, s\, m_{\sss Z}^2\, \epsilon^3\, I_{1,1,0,1,2,0,1}
                             $}
\nn
&
&
                      &- 4\, s\, m_q^2\, \epsilon^2\, I_{1,1,0,2,2,0,0} \,,
\end{align}
}
where $\lambda_{1,2,3}$ are three square roots related to kinematic variables,
\begin{equation}
\label{eq:lambda}
\lambda_1 = \sqrt{s\, ( s - 4\, m_q^2 )}\,,
\qquad
\lambda_2 = \sqrt{m_{\sss Z}^2\, ( m_{\sss Z}^2 - 4\, m_q^2 )}\,,
\qquad
\lambda_3 = \sqrt{\vpq{s\, ( s - 4\, m_{\sss Z}^2 )}}\,.
\end{equation}
The canonical basis $\mathbf{g}(\vec{x},\epsilon)$ defined in Eq.\eqref{eq:UTbasis} is dimensionless and can be expressed as a Taylor series in $\epsilon$. For the two-loop three-point integral family with three distinct off-shell legs and a massive internal loop, four square roots emerge that cannot be simultaneously rationalized \cite{Wang:2019fxh,Chaubey:2022hlr}. However, in our specific case, these four square roots degenerate into the three square roots defined in Eq.\eqref{eq:lambda}. Fortunately, the square roots $\lambda_{1,2,3}$ stemming from the NNLO $\text{QCD} \otimes \text{EW}$ corrections to $e^+e^- \rightarrow ZZ$ can be simultaneously  rationalized as
\begin{align}
\frac{\lambda_1}{m_q^2}
&=
\frac{\big[\, (1-y)^2 + x\, y\, (2-x) \,\big]\,
\big[\, (1-y)^2 + x^2\, y \,\big]}
{x\, y\, (1-x)\, \big[\, (1-y)^2 + x\, y \,\big]}
\nonumber \\
\frac{\lambda_2}{m_q^2}
&=
\frac{(1-y)\,(1+y)}{y}
\\
\frac{\lambda_3}{m_q^2}
&=
\frac{\big[\, (1-y)^2 + x^2\, y \,\big]\,
\big[\, (2\, x - 1)\, (1 - y)^2 + x^2\, y \,\big]}{x\, y\, (1-x)\,
\big[\, (1-y)^2 + x\, y \,\big]}
\nonumber
\end{align}
by the following change of variables,
\begin{equation}
-\frac{s}{m_q^2}
=
\frac{\big[\, (1-y)^2+x^2\, y \,\big]^2}
{x\, y\, (1-x)\, \big[\, (1-y)^2 + x\, y \,\big]}\,,
\qquad\quad
-\, \frac{m_{\sss Z}^2}{m_q^2}
=
\frac{(1-y)^2}{y}\,.
\end{equation}
Consequently, the canonical differential equations for $\mathbf{g}(x,y,\epsilon)$ can be cast into the following $d \log$ form,
\begin{equation}
d{\mathbf{g}}(x,y,\epsilon)
=
\epsilon\,
\Big[\sum_{i=1}^{18} \mathbb{M}_i \dlog \eta_{i}(x,y) \Big]\,
\mathbf{g}(x,y,\epsilon)\,,
\end{equation}
where the $18$ rational symbol letters are given as
\begin{align}
\label{eq:alphabet}
\eta_{1}  &= x \,, &
\eta_{10} &= 1 - y\,(1-x\,y) \,,
\nn
\eta_{2}  &= y \,, &
\eta_{11} &= (1-y)^2 + x\, y \,,
\nn
\eta_{3}  &= 1 - x \,, &
\eta_{12} &= (1-y)^2 + x^{2}\, y \,,
\nn
\eta_{4}  &= 1 - y \,, &
\eta_{13} &= (1-y)\,(x-y) + x^2\, y \,,
\nn
\eta_{5}  &= 1 + y \,, &
\eta_{14} &= (1-y)^2 + x\, y\,(2-x) \,,
\\
\eta_{6}  &= 1 - x - y \,, &
\eta_{15} &= (1-y)\,(1-x\, y) + x^{2}\, y \,,
\nn
\eta_{7}  &= 1 - y\,(1-x) \,, &
\eta_{16} &= (2\,x-1)\,(1-y)^2 + x^{2}\, y \,,
\nn
\eta_{8}  &= 1 - y\,(1-y) \,, &
\eta_{17} &= (1-x-y)\,(1-y)^2 - x^2\,y^2 \,,
\nn
\eta_{9}  &= x - y\,(1-y) \,, &
\eta_{18} &= (1-y+x\, y)\,(1-y)^2 + x^{2}\,y \,,
\nonumber
\end{align}
and the explicit expressions of the constant matrices $\mathbb{M}_i$ are available in the supplementary file ``dlog-fom\_Matrix.m." In the Euclidean region delineated by
\begin{equation}
\frac{1}{2} < x < 1-2\, y
\quad \land \quad
0 < y < \frac{1}{4}\,,
\end{equation}
all letters are real and positive, and thus all MIs are real functions of the dimensionless variables $x$ and $y$. Furthermore, it is evident that the differential one-form $\dA$ is composed exclusively of rational functions, as all the symbol letters in Eq.\eqref{eq:alphabet} are rational functions of the kinematic variables. Therefore, the iterated integral solution \eqref{eq:pathorderedexp} of this canonical differential system can be expressed in terms of GPLs. The GPLs are recursively defined by \cite{Goncharov:1998kja}
\begin{equation}
G(a_1, \dots, a_n; z)
=
\int_0^z \frac{1}{t-a_1}\,G(a_{2}, \dots, a_n;t)\, {d}t
\end{equation}
with
\begin{equation}
G(\,; z) = 0
\qquad
\text{and}
\qquad
G(\underbrace{0,\dots,0}_{n \text{ times}};z)=\frac{\log^n(z)}{n!}\,,
\end{equation}
where $(a_1,\ldots, a_n)$ is referred to as the weight vector of $G(a_1, \dots, a_n; z)$, and the length of the weight vector is called the weight of the GPL.

\par
Within the framework of the differential system constructed using the canonical basis as outlined in Eq.\eqref{eq:UTbasis}, we determine the integration constants by an independent and simpler calculation of the canonical basis at the boundary point $\vec{x}_0 = (0, \,1)$, which corresponds to the vanishing external momenta $s = m_{\sss Z}^2 = 0$. At this specific kinematic point, all MIs in the pre-canonical basis ${\mathbf{I}}(\vec{x},\epsilon)$ degenerate into equal-mass vacuum integrals. Additionally, it is noteworthy that all canonical MIs $g_i\, (i=1,\dots,26)$ are regular at the boundary point $\vec{x}_0$. Through a straightforward analysis of the canonical transformation in Eq.\eqref{eq:UTbasis}, we conclude that all canonical MIs vanish at $\vec{x}_0$ except for $g_1$. Under the normalization of the integration measure specified in Eq.\eqref{eq:measure}, the boundary conditions are formulated as
\begin{equation}
g_i(\vec{x}_0,\epsilon) = \left\{
\begin{matrix}
& 1 \,, \qquad & \quad i = 1
\\
& 0 \,, \qquad & \quad i \neq 1
\end{matrix}
\right.
\end{equation}
Leveraging these boundary conditions, we are able to derive the analytic solution of the canonical differential system. All canonical MIs are expressed as Taylor series in $\epsilon$, with coefficients articulated in terms of GPLs. The analytic expressions of these canonical MIs, $g_i~ (i=1, \dots, 26)$, are presented in Appendix \ref{sec:GPLs} up to the order of $\epsilon^2$. For additional convenience and accessibility, the complete analytic expressions for all MIs up to $\mathcal{O}(\epsilon^4)$ are available in the supplementary file ``analytic\_MIs.m," which accompanies the arXiv submission of this paper.

\subsection{\label{sec:numcheck}Numerical checks}

\par
We utilize the \texttt{Mathematica} package \texttt{PolyLogTools} \cite{Maitre:2005uu,Maitre:2007kp,Duhr:2019tlz} and the \texttt{C++}  library \texttt{GiNaC} \cite{Bauer:2000cp,Vollinga:2004sn} for the symbolic computation and numerical evaluation of GPLs. To ensure the accuracy and reliability of our analytic results, we also perform numerical checks on the $26$ MIs defined by Eq.\eqref{eq:UTbasis} against the numerical results from the publicly available package \texttt{AMFlow}. We find excellent agreement between our analytical and numerical results, with an accuracy exceeding $10^{-99}$, i.e.,
\begin{equation}
\left|
1 -  
g_i^{{\sss \mathrm{(AMF)}}}/g_i^{{\sss \mathrm{(GPL)}}}
\right|
<
10^{-99}
\qquad
(i=1, \dots, 26)\,,
\end{equation}
where $g_i^{{\sss \mathrm{(GPL)}}}$ and $g_i^{{\sss \mathrm{(AMF)}}}$ denote the numerical results obtained from our analytic expressions and those obtained using the \texttt{AMFlow} package, respectively. In Table \ref{table1}, we present the numerical values for two representative MIs, $g_{23}$ and $g_{26}$, evaluated at the Euclidean point $(x, y)=(0.5, 0.2)$ up to $\mathcal{O}(\epsilon^4)$, accurate to $90$ decimal places. The complete numerical results for all $g_i~ (i = 1, \ldots, 26)$ at this validation point with 100-digit precision are provided in the supplementary files ``numMIs\_GPL.m" and ``numMIs\_AMF.m," respectively.
\begin{table}[htbp]
\centering
\renewcommand{\arraystretch}{1.1}
\resizebox{1.00\linewidth}{!}{
\begin{tabular}{|ccc|
}
\toprule[1.2pt]
\multirow{1}{*}{\quad{\bf MI}\quad} 
&
\multirow{1}{*}{\bf Order}
&
\multirow{1}{*}{{\bf Numerical value}}
\\
\midrule[0.4pt]
\midrule[0.4pt]
\multirow{5}{*}{$g_{23}$} & {$\epsilon^0$}
&
{$0$}
\\
& {$\epsilon^1$}
& {$0$}
\\
& {$\epsilon^2$}
& {$-\,7.26101564334485215986593566256166202846742579354269938897134626597079498884306857349066237$\quad}
\\
& {$\epsilon^3$}
& {$~\,11.86463522232032158447131561240614580110409186179734960547391102577769656727839159997486514$\quad}
\\
& {$\epsilon^4$}
& {$-\,9.28510545026443974881656592262564288473103903999133306919325410029473865878053069212244617$\quad}
\\
\midrule[0.4pt]
\multirow{5}{*}{$g_{26}$} & {$\epsilon^0$}   & {$0$}
\\
& {$\epsilon^1$}
& {$0$}
\\
& {$\epsilon^2$}
& {$\quad7.26101564334485215986593566256166202846742579354269938897134626597079498884306857349066237$\quad}
\\
& {$\epsilon^3$}
& {$-\,9.68816618676192114865212264256976565734577230132356179420044708343871148651833620502830076$\quad}
\\
& {$\epsilon^4$}
& {$\quad3.91185381115700898808123227066867633509446778857243575613151816779834900833108054340221357$\quad}
\\
\bottomrule[1.2pt]
\end{tabular}
}
\caption{Numerical values for two representative MIs from the sectors $\mathcal{I}_{13467}$ and $\mathcal{I}_{12457}$, evaluated at $(x, y) = (0.5, 0.2)$ up to $\mathcal{O}(\epsilon^4)$, with precision to $90$ decimal places.}
\label{table1}
\end{table}

\section{\label{sec:4}Phenomenological results and discussion}

\par
In this section, we calculate the integrated cross section and various kinematic distributions for $e^+ e^- \rightarrow ZZ$ using the analytic expressions of the MIs derived in Sec.\ref{sec:3}, and present a detailed phenomenological analysis based on our calculations. We neglect the masses of all fermions except for the top and bottom quarks. All relevant SM input parameters are taken as follows \cite{ParticleDataGroup:2022pth}:
\begin{align}
&m_{\sss Z} = 91.1876~\mathrm{GeV}\,, &
&m_t = 172.5~\mathrm{GeV}\,, &
&\alpha(0)  = 1/137.035999084\,,
\nonumber \\
&m_{\sss W} = 80.377~\mathrm{GeV}\,, &
&m_b = 4.78~\mathrm{GeV}\,, &
&G_{\mu} = 1.1663787\times 10^{-5}~\mathrm{GeV}^{-2}\,,
\\
&m_{\sss H} = 125.25~\mathrm{GeV}\,, &
&\alpha_s(m_{\sss Z}) = 0.1179\,. &
\nonumber
\end{align}
The renormalization scale for the strong coupling constant is chosen as $\mu = m_{\sss Z}$. For comparison purposes, the numerical computations are performed in both the $\alpha(0)$ and $G_{\mu}$ schemes.

\subsection{\label{sec:4A}Integrated cross sections}

\par
\begin{figure}[htbp]
\centering
\includegraphics[width = 1.0\textwidth]{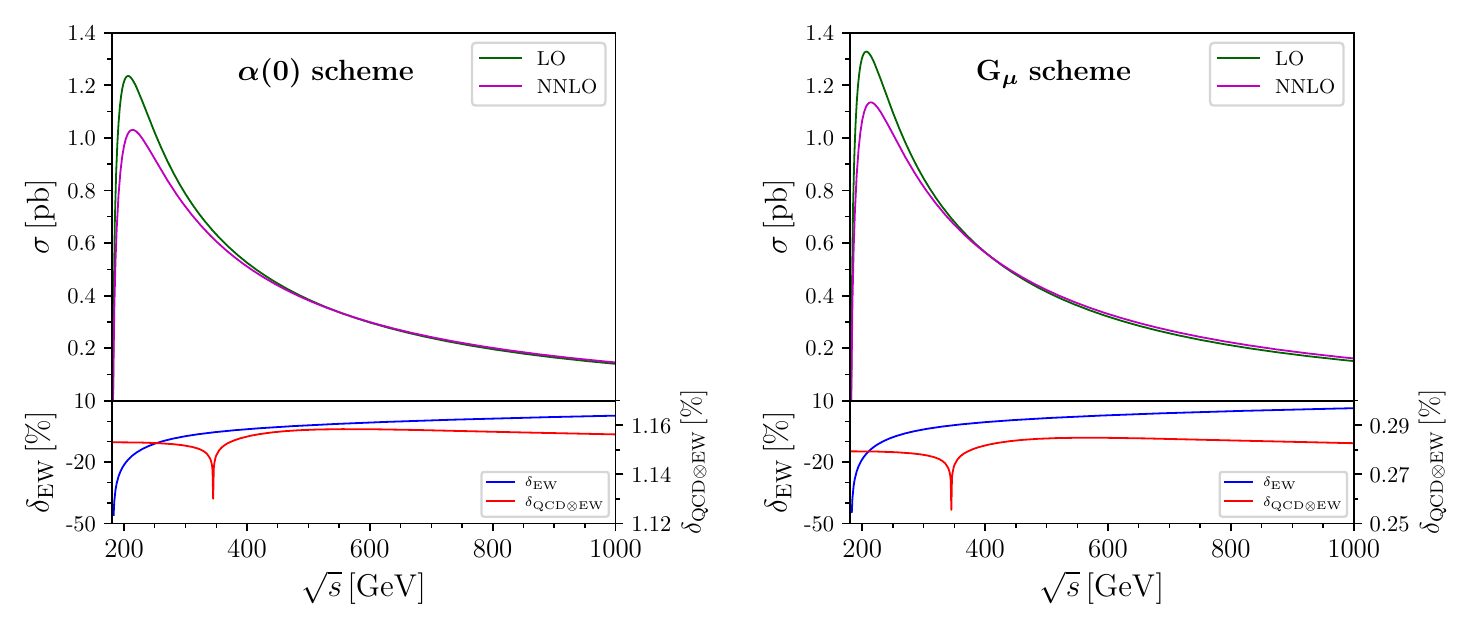}
\caption{LO and NNLO corrected integrated cross sections, along with the corresponding EW and $\text{QCD} \otimes \text{EW}$ relative corrections, for $e^+e^- \rightarrow ZZ$ as functions of the $e^+e^-$ colliding energy in both the $\alpha(0)$ (left) and $G_{\mu}$ (right) schemes.}
\label{fig3}
\end{figure}

To facilitate the presentation of the integrated cross section for $e^+ e^- \rightarrow ZZ$, incorporating both the NLO EW corrections and the NNLO mixed $\text{QCD} \otimes \text{EW}$ corrections, we express the cross section as
\begin{equation}
\sigma_{\sss \mathrm{NNLO}}
=
\sigma_{\sss \mathrm{LO}}\, (1+\delta_{\sss \mathrm{EW}} + \delta_{\sss \mathrm{QCD \otimes EW}})\,,
\end{equation}
where $\delta_{\sss \mathrm{EW}}$ and $\delta_{\sss \mathrm{QCD \otimes EW}}$ denote the relative corrections for the NLO EW and NNLO mixed $\text{QCD} \otimes \text{EW}$ contributions, respectively, and are defined by
\begin{equation}
\delta_{\sss \mathrm{EW}}
=
\frac{\Delta\sigma_{\sss \mathrm{EW}} }{\sigma_{\sss \mathrm{LO}}}\,,
\qquad\quad
\delta_{\sss \mathrm{QCD \otimes EW}}
=
\frac{\Delta\sigma_{\sss \mathrm{QCD \otimes EW}} }{\sigma_{\sss \mathrm{LO}}}\,.
\end{equation}
The LO and NNLO corrected integrated cross sections for the process $e^+ e^- \rightarrow ZZ$, as functions of the $e^+e^-$ colliding energy $\sqrt{s}$, are depicted in Fig.\ref{fig3} for both the $\alpha(0)$ scheme (left) and the $G_{\mu}$ scheme (right). The corresponding EW and $\text{QCD} \otimes \text{EW}$ relative corrections are presented in the lower panels. As illustrated in this figure, the production cross sections exhibit similar dependencies on $\sqrt{s}$ in both schemes. The LO integrated cross sections are sensitive to the colliding energy, increasing sharply near the $Z$-pair production threshold as $\sqrt{s}$ rises, peaking around $\sqrt{s}\sim 207~\mathrm{GeV}$, and then decreasing smoothly at higher energies. As shown in the lower panels of Fig.\ref{fig3}, the EW correction significantly suppresses the LO cross section near the threshold region, which can be attributed to the Coulomb singularity effect \cite{Denner:1991kt,Beenakker:1996kt,Chen:2014iwb,Zhang:2015prr}. The NLO EW relative correction increases from approximately $-45\%$ to around $2.7\%$ in the $\alpha(0)$ scheme and to around $6.3\%$ in the $G_{\mu}$ scheme as $\sqrt{s}$ increases from the threshold to $1000~\mathrm{GeV}$. However, the $\text{QCD} \otimes \text{EW}$ correction enhances the LO cross section across the entire plotted $\sqrt{s}$ region. The relative correction exceeds $1.1\%$ in the $\alpha(0)$ scheme and amounts to approximately $0.3\%$ in the $G_{\mu}$ scheme. Compared to the $\text{QCD} \otimes \text{EW}$ correction in the $\alpha(0)$ scheme, the $\text{QCD} \otimes \text{EW}$ correction in the $G_{\mu}$ scheme is reduced by about an order of magnitude. This effect is attributed to the incorporation of certain significant higher-order corrections into the LO cross section \cite{Denner:1991kt,Beenakker:1996kt,Denner:2019vbn}. In the $\text{QCD} \otimes \text{EW}$ relative correction, a resonance peak appears at $\sqrt{s}=2\, m_{t}$ due to the resonance effect caused by top-quark loop integrals at the pseudo-threshold of top-quark pair production. To present the numerical results more precisely, we summarize in Table \ref{tab2} the LO, NLO and NNLO integrated cross sections, as well as the corresponding EW and $\text{QCD} \otimes \text{EW}$ relative corrections, for some representative colliding energies in both the $\alpha(0)$ and $G_{\mu}$ schemes.
\begin{table}[htbp]
\centering
\renewcommand{\arraystretch}{1.0}
\begin{tabular}{
p{1.8cm}<{\centering}
p{1.6cm}<{\centering}
p{2.4cm}<{\centering}
p{2.4cm}<{\centering}
p{1.8cm}<{\raggedleft}
p{2.8cm}<{\centering}
p{2.0cm}<{\centering}
}
\toprule[1.2pt]
\multirow{1}{*}{$\sqrt{s}~ \text{[GeV]}$ }
&
\multirow{1}{*}{Scheme}
&
\multirow{1}{*}{$\sigma_{\sss{\text{LO}}}~ \text{[pb]}$}
&
\multirow{1}{*}{$\sigma_{\sss{\text{NLO}}}~ \text{[pb]}$}
&
\multirow{1}{*}{$\delta_{\sss{\text{EW}}}~ \text{[\%]}~~$}
&
\multirow{1}{*}{$\sigma_{\sss{\text{NNLO}}}~ \text{[pb]}$}
&
\multirow{1}{*}{$\delta_{\sss{\text{QCD} \otimes \text{EW}}}~ \text{[\%]}$}
\\
\midrule[0.4pt]
\midrule[0.4pt]
\multirow{2}{*}{$240$}
& {$\alpha(0)$}
& {$1.077615$}
& {$0.94682$}
& {$-12.138~~$}
& {$0.95924$}
& {$1.1529$}
\\
& {$G_{\mu}$}
& {$1.158461$}
& {$1.05435$}
& {$-8.987~~$}
& {$1.05759$}
& {$0.2792$}
\\
\midrule[0.4pt]
\multirow{2}{*}{$250$}
& {$\alpha(0)$}
& {$1.019355$}
& {$0.90740$}
& {$-10.983~~$}
& {$0.91915$}
& {$1.1528$}
\\
& {$G_{\mu}$}
& {$1.095830$}
& {$1.01048$}
& {$-7.789~~$}
& {$1.01354$}
& {$0.2791$}
\\
\midrule[0.4pt]
\multirow{2}{*}{$350$}
& {$\alpha(0)$}
& {$0.632563$}
& {$0.59909$}
& {$-5.291~~$}
& {$0.60635$}
& {$1.1477$}
\\
& {$G_{\mu}$}
& {$0.680020$}
& {$0.66712$}
& {$-1.897~~$}
& {$0.66898$}
& {$0.2738$}
\\
\midrule[0.4pt]
\multirow{2}{*}{$500$}
& {$\alpha(0)$}
& {$0.385130$}
& {$0.37745$}
& {$-1.994~~$}
& {$0.38191$}
& {$1.1581$}
\\
& {$G_{\mu}$}
& {$0.414023$}
& {$0.42026$}
& {$1.507~~$}
& {$0.42144$}
& {$0.2848$}
\\
\midrule[0.4pt]
\multirow{2}{*}{$1000$}
& {$\alpha(0)$}
& {$0.140378$}
& {$0.14419$}
& {$2.721~~$}
& {$0.14582$}
& {$1.1562$}
\\
& {$G_{\mu}$}
& {$0.150909$}
& {$0.16051$}
& {$6.368~~$}
& {$0.16094$}
& {$0.2830$}
\\
\bottomrule[1.2pt]
\end{tabular}
\caption{LO, NLO and NNLO corrected integrated cross sections, as well as the corresponding EW and $\text{QCD} \otimes \text{EW}$ relative corrections, for $e^+e^- \rightarrow ZZ$ at some representative colliding energies in both the $\alpha(0)$ and $G_{\mu}$ schemes.}
\label{tab2}
\end{table}

\subsection{\label{sec:4B}Kinematic distributions}

\par
In this subsection, we analyze the kinematic distributions of the final-state $Z$ boson for $e^+e^- \rightarrow ZZ$ at $\sqrt{s} = 240~\mathrm{GeV}$. The EW and $\text{QCD} \otimes \text{EW}$ differential relative corrections with respect to the kinematic variable $x$ are defined as
\begin{equation}
\delta_{\sss \mathrm{EW}}
=
\Big(
\frac{\mathrm{d}\sigma_{\sss \mathrm{EW}}}{\mathrm{d}x}
-
\frac{\mathrm{d}\sigma_{\sss \mathrm{LO}}}{\mathrm{d}x}
\Big)
\Big/
\frac{\mathrm{d}\sigma_{\sss \mathrm{LO}}}{\mathrm{d}x}\,,
\qquad\quad
\delta_{\sss \mathrm{QCD \otimes EW}}
=
\Big(
\frac{\mathrm{d}\sigma_{\sss \mathrm{QCD \otimes EW}}}{\mathrm{d}x}
-
\frac{\mathrm{d}\sigma_{\sss \mathrm{LO}}}{\mathrm{d}x}
\Big)
\Big/
\frac{\mathrm{d}\sigma_{\sss \mathrm{LO}}}{\mathrm{d}x}\,.
\end{equation}
Given the indistinguishability of the two $Z$ bosons in the final state, the kinematic distributions of the final-state $Z$ boson are defined as the average of the distributions for the two identical $Z$ bosons.

\par
\begin{figure}[htbp]
\centering
\includegraphics[width = 1.0\textwidth]{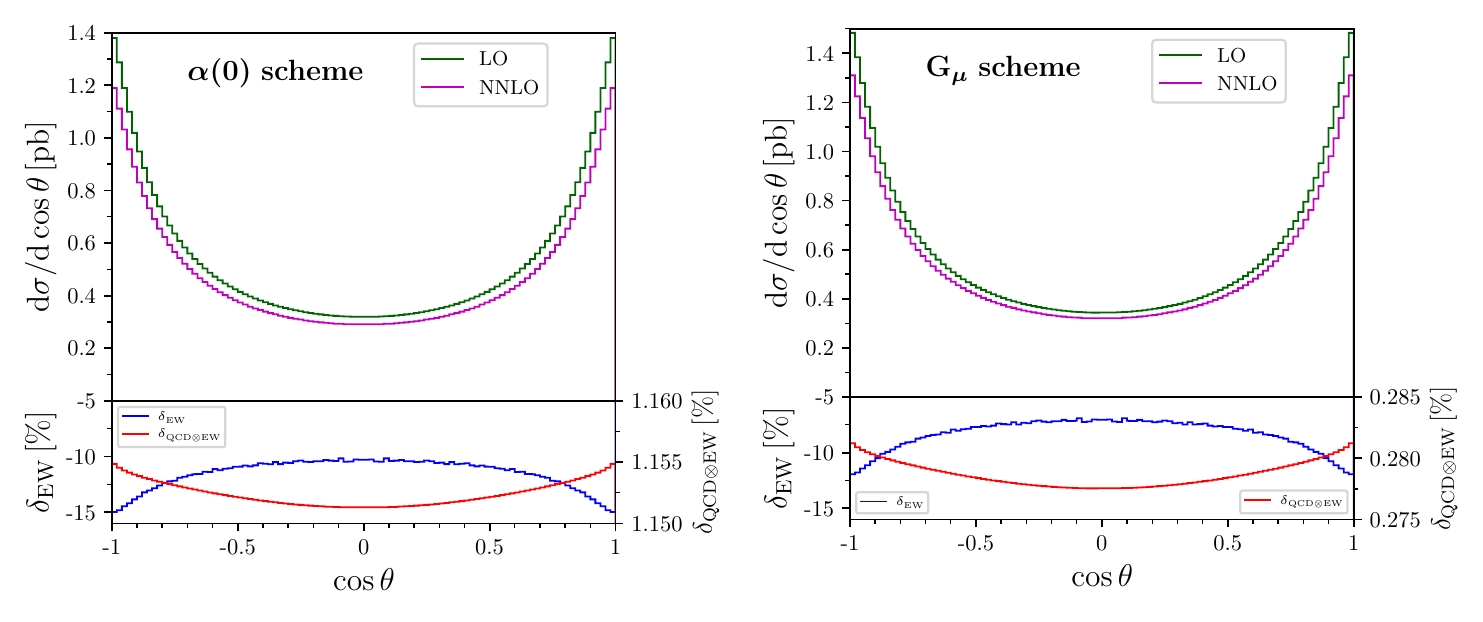}
\caption{LO and NNLO corrected scattering angle distributions of the final-state $Z$ boson, along with the corresponding EW and $\text{QCD} \otimes \text{EW}$ relative corrections, for $e^+ e^- \rightarrow ZZ$ at $\sqrt{s } = 240~\mathrm{GeV}$ in both the $\alpha(0)$ (left) and $G_{\mu}$ (right) schemes.}
\label{fig4}
\end{figure}

Due to $\mathcal{CP}$ conservation and Bose symmetry, the scattering angle distribution of the final-state $Z$ boson exhibits forward-backward symmetry, i.e., ${d\sigma}/{d\cos\theta}$ is symmetric with respect to $\cos\theta = 0$. In Fig.\ref{fig4}, we illustrate the LO and NNLO corrected scattering angle distributions of final-state $Z$ boson, as well as the corresponding EW and $\text{QCD} \otimes \text{EW}$ relative corrections, for $e^+e^- \rightarrow ZZ$ at $\sqrt{s} = 240~\mathrm{GeV}$ in both the $\alpha(0)$ and $G_{\mu}$ schemes. As shown in the figure, the scattering angle distribution exhibits strong peaks in both the forward and backward directions,  indicating that $Z$-boson pairs are predominantly produced along the directions of the electron and positron beams. The EW correction suppresses the LO differential cross section across the entire range of $\cos\theta$. The corresponding relative correction is approximately $-10\%$ in the $\alpha(0)$ scheme and approximately $-7\%$ in the $G_{\mu}$ scheme at $\cos\theta = 0$,  decreasing by about $5\%$ as $|\cos\theta|$ increases to $1$. In contrast, the $\text{QCD} \otimes \text{EW}$ correction slightly enhances the LO differential cross section, with specific increases of approximately $1.15\%$ in the $\alpha(0)$ scheme and about $0.28\%$ in the $G_{\mu}$ scheme, particularly in the forward and backward directions. Notably, the forward-backward symmetry in the scattering angle distribution of the final-state $Z$ boson is clearly shown in Fig.\ref{fig4}.

\par
\begin{figure}[htbp]
\centering
\includegraphics[width = 1.0\textwidth]{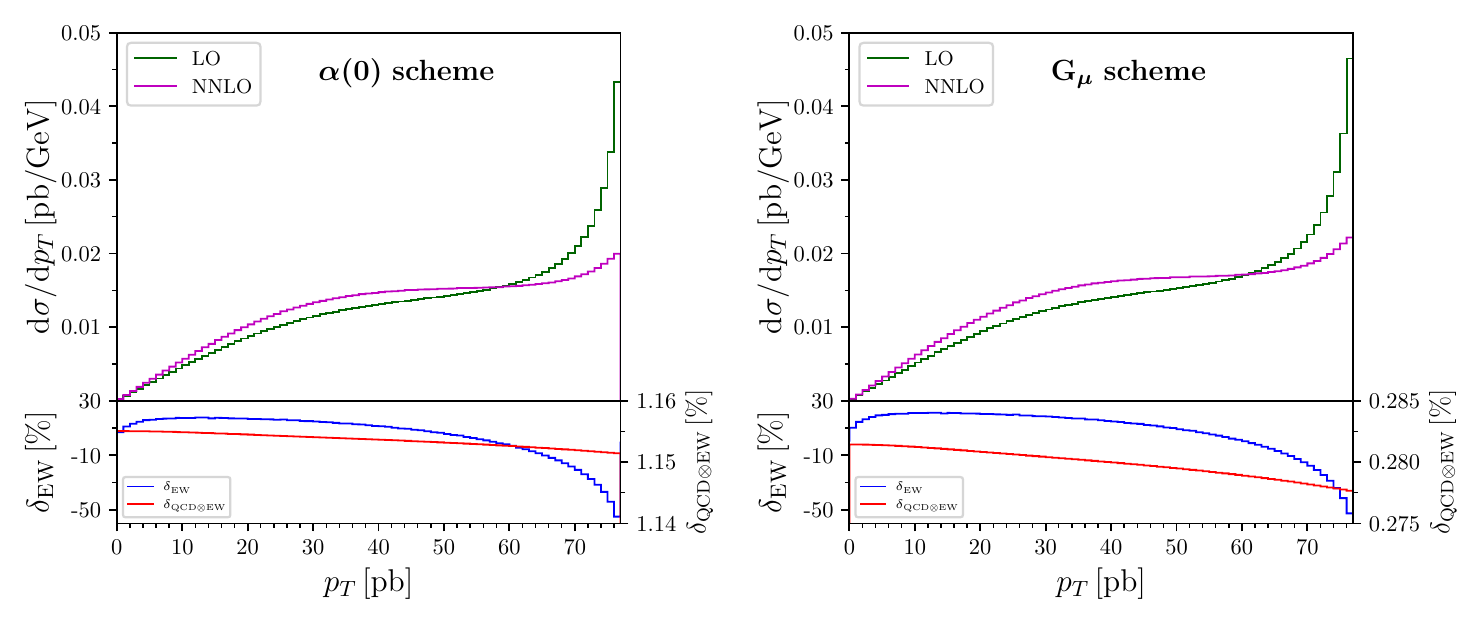}
\caption{Same as Fig.\ref{fig4}, but for transverse momentum distributions of the final-state $Z$ boson.}
\label{fig5}
\end{figure}

The LO and NNLO corrected transverse momentum distributions of the final-state $Z$ boson, as well as the corresponding EW and $\text{QCD} \otimes \text{EW}$ relative corrections, at $\sqrt{s} = 240~\mathrm{GeV}$ in both the $\alpha(0)$ and $G_{\mu}$ schemes, are presented in Fig.\ref{fig5}. The LO differential cross section increases smoothly with increasing $p_T$ in the low $p_T$ region, while it rises sharply with $p_T$ in the high $p_T$ region. The EW correction provides a moderate enhancement to the LO differential cross section in the low $p_T$ region, while suppressing it in the high $p_T$ region, where the EW relative correction decreases rapidly as $p_T$ increases. The substantial magnitude of the EW relative correction at extremely high $p_T$ can be attributed to the Sudakov effect. Conversely, the $\text{QCD} \otimes \text{EW}$ correction results in a slight enhancement of the LO differential cross section across the entire $p_T$ region. The relative correction decreases very slowly, remaining steady at around $1.15\%$ in the $\alpha(0)$ and around $0.28\%$ in the $G_{\mu}$ scheme. The Sudakov effect is not observed in the NNLO $\text{QCD} \otimes \text{EW}$ correction because there are no real emission corrections at the $\alpha\alpha_s$ order.

\par
Based on the discussion of the scattering angle and transverse momentum distributions, we conclude that the NNLO $\text{QCD} \otimes \text{EW}$ corrections will have a significant impact on the interpretation of future high-precision experimental measurements. In the $\alpha(0)$ scheme, the $\text{QCD} \otimes \text{EW}$ relative correction exceeds $1.15\%$, while in the $G_{\mu}$ scheme, it amounts to approximately $0.28\%$ across most of the phase space. Near the top-pair production threshold, the $\text{QCD} \otimes \text{EW}$ relative correction is marginally reduced, yet still surpassing approximately $1.12\%$ and $0.25\%$ in the $\alpha(0)$ and $G_{\mu}$ schemes, respectively.

\section{\label{sec:5}Summary}

\par
The production of $Z$-boson pair serves as a crucial avenue for comprehending the EW gauge structure, investigating anomalous TGCs, and searching for new physics. Inspired by the anticipated high-precision measurements and the cleaner hadronic environment of future lepton colliders, we present a detailed calculation of the NNLO $\text{QCD} \otimes \text{EW}$ corrections to $Z$-boson pair production at electron-positron colliders and provide a more precise theoretical prediction in this paper. Using the canonical differential equation method, we perform a comprehensive analytic calculation of all two-loop MIs arising from the $\mathcal{O}(\alpha\alpha_s)$ corrections. In particular, we derive the canonical MIs for the massive two-loop triangle MIs and express them in terms of GPLs up to the order of $\epsilon^4$. The analytic expressions offer a faster and more accurate numerical evaluation for any kinematic configuration compared to purely numerical methods such as sector decomposition, which is essential for precise theoretical predictions and phenomenological analysis.

\par
We apply our analytic expressions for these two-loop canonical MIs to compute the NNLO $\text{QCD} \otimes \text{EW}$ corrected integrated cross sections for $e^+e^- \rightarrow ZZ$, as well as the differential distributions with respect to the scattering angle and transverse momentum of the final-state $Z$ boson. Our findings indicate that the NNLO $\text{QCD} \otimes \text{EW}$ corrections are non-negligible, leading to an enhancement of the LO cross section across the entire phase space. In the $\alpha(0)$ scheme, the $\text{QCD} \otimes \text{EW}$ relative correction reaches approximately $1.15\%$, whereas in the $G_{\mu}$ scheme, it is reduced to about $0.28\%$. This seems to indicate that the perturbative convergence in the $G_{\mu}$ scheme is superior to that in the $\alpha(0)$ scheme. To align with the anticipated permille level of accuracy in experimental measurements at future lepton colliders, these NNLO $\text{QCD} \otimes \text{EW}$ corrections should be incorporated into the theoretical predictions.

\vskip 8mm
\noindent{\large\bf Acknowledgments:}
\par
This work is supported by the National Natural Science Foundation of China (Grant No. 12061141005) and the CAS Center for Excellence in Particle Physics (CCEPP).
Zhe Li would like to express heartfelt gratitude to Ming-Ming Long for collaboration in the early stages of the project.

\appendix
\section{\label{sec:GPLs}Explicit expressions of canonical MIs}
In this appendix, we present the explicit analytic expressions of $g_{i}~ (i = 1, \dots, 26)$ in terms of GPLs up to $\mathcal{O}(\epsilon^2)$ as follows:
{
\fontsize{9.5pt}{15}\selectfont
\begin{align}
g_{1} = {} &
            1 \,,
\nn
g_{2} = {} &
            \epsilon\,
            \big[\,
            G(0; y)
            - 2\, G(1; y)
            + G(0; x)
            + G(1; x)
            - G(-{(1-y)^{2}}/{y}; x)
            \,\big]
\nn
&
            + \epsilon^{2}\,
            \big[\,
            G(0; x)\, G(0; y)
            + G(1; x)\, G(0; y)
            - 2\, G(0; x)\, G(1; y)
            - 2\, G(1; x)\, G(1; y)
            - G(1, -{(1-y)^{2}}/{y}; x)
\nn
&
            + G(0; y)\, G(-{(1-y)^{2}}/{y}; x)
            - 2\, G(0; y)\, G(1-\sqrt{y+{1}/{y}-1}; x)
            - 2\, G(0; y)\, G(1+\sqrt{y+{1}/{y}-1}; x)
\nn
&
            - 2\, G(1; y)\, G(-{(1-y)^{2}}/{y}; x)
            + 4\, G(1; y)\, G(1-\sqrt{y+{1}/{y}-1}; x)
            + 4\, G(1; y)\, G(1+\sqrt{y+{1}/{y}-1}; x)
\nn
&
            - 2\, G(1-\sqrt{y+{1}/{y}-1}, 0; x)
            - 2\, G(1-\sqrt{y+{1}/{y}-1}, 1; x)
            + 2\, G(1-\sqrt{y+{1}/{y}-1}, -{(1-y)^{2}}/{y}; x)
\nn
&
            - 2\, G(1+\sqrt{y+{1}/{y}-1}, 0; x)
            - 2\, G(1+\sqrt{y+{1}/{y}-1}, 1; x)
            + 2\, G(1+\sqrt{y+{1}/{y}-1}, -{(1-y)^{2}}/{y}; x)
\nn
&
            - G(0, -{(1-y)^{2}}/{y}; x)
            + G(-{(1-y)^{2}}/{y}, 0; x)
            + G(-{(1-y)^{2}}/{y}, 1; x)
            - 2\, G(0, 1; y)-2\, G(1, 0; y)
\nn
&
            - G(-{(1-y)^{2}}/{y}, -{(1-y)^{2}}/{y}; x)
            + G(0, 0; x)
            + G(0, 1; x)
            + G(1, 0; x)
            + G(1, 1; x)
            + 4\, G(1, 1; y)
\nn
&
            + G(0, 0; y)-{\pi ^{2}}/{6}
            \,\big]
            + \mathcal{O}(\epsilon^{3}) \,,
\nn
g_{3} = {} &
            \epsilon\, \, G(0; y)
            + \epsilon^{2}\,
            \big[\,
            G(0, 0; y)
            - 2\, G(-1, 0; y)
            - {\pi ^{2}}/{6}
            \,\big]
            + \mathcal{O}(\epsilon^{3}) \,,
\nn
g_{4} = {} &
            2\, \epsilon^{2}\,
            \big[\,
            G(0; x)\, G(0; y)
            + G(1; x)\, G(0; y)
            - 2\, G(0; x)\, G(1; y)
            - 2\, G(1; x)\, G(1; y)
            - G(0; y)\, G(-{(1-y)^{2}}/{y}; x)
\nn
&
            + 2\, G(1; y)\, G(-{(1-y)^{2}}/{y}; x)
            + G(-{(1-y)^{2}}/{y}, -{(1-y)^{2}}/{y}; x)
            - G(0, -{(1-y)^{2}}/{y}; x)
            - 2\, G(0, 1; y)
\nn
&
            - G(1, -{(1-y)^{2}}/{y}; x)
            - G(-{(1-y)^{2}}/{y}, 0; x)
            - G(-{(1-y)^{2}}/{y}, 1; x)
            + 4\, G(1, 1; y)
            - 2\, G(1, 0; y)
\nn
&
            + G(0, 0; x)
            + G(0, 1; x)
            + G(1, 0; x)
            + G(1, 1; x)
            + G(0, 0; y)
            \,\big]
            + \mathcal{O}(\epsilon^{3}) \,,
\nn
g_{5} = {} &
            \epsilon^{2}\,
            \big[\,
            G(0; x)\, G(0; y)
            + G(1; x)\, G(0; y)
            - G(0; y)\, G(-{(1-y)^{2}}/{y}; x)
            + 2\, G(0, 0; y)
            - 2\, G(0, 1; y)
\nn
&
            - 2\, G(1, 0; y)
            \,\big]
            + \mathcal{O}(\epsilon^{3}) \,,
\nn
g_{6} = {} &
            2\, \epsilon^{2}\, G(0, 0; y)
            + \mathcal{O}(\epsilon^{3}) \,,
\nn
g_{7} = {} &
            \epsilon^{2}\,
            \big[\,
            G(0; x)\, G(0; y)
            - G(1; x)\, G(0; y)
            - 2\, G(0; x)\, G(1; y)
            + 2\, G(1; x)\, G(1; y)
            + G(0; y)\, G(1-y; x)
\nn
&
            + 2\, G(1; y)\, G(1-y; x)
            - G(0; y)\, G((1-y) y; x)
            + 2\, G(1; y)\, G({(y-1)}/{y}; x)
            - 3\, G(0; y)\, G({(y-1)}/{y}; x)
\nn
&
            + 3\, G(0; y)\, G({(y-1)}/{y^{2}}; x)
            - 2\, G(1; y)\, G({(y-1)}/{y^{2}}; x)
            - G(0; y)\, G(-{(1-y)^{2}}/{y}; x)
            + 4\, G(1, 1; y)
\nn
&
            + 2\, G(1; y)\, G(-{(1-y)^{2}}/{y}; x)
            - 2\, G(1; y)\, G(-((y-1) y); x)
            + G({(y-1)}/{y^{2}}, 1; x)
            + G({(y-1)}/{y^{2}}, 0; x)
\nn
&
            - G({(y-1)}/{y^{2}}, -{(1-y)^{2}}/{y}; x)
            - G(0, -{(1-y)^{2}}/{y}; x)
            + G(1, -{(1-y)^{2}}/{y}; x)
            - G(-{(1-y)^{2}}/{y}, 0; x)
\nn
&
            + G(1-y, -{(1-y)^{2}}/{y}; x)
            + G({(y-1)}/{y}, -{(1-y)^{2}}/{y}; x)
            - G({(y-1)}/{y}, 1; x)
            - G(-{(1-y)^{2}}/{y}, 1; x)
\nn
&
            + G(-{(1-y)^{2}}/{y}, -{(1-y)^{2}}/{y}; x)
            + G(-((y-1) y), 0; x)+ G(-((y-1) y), 1; x)
            - G({(y-1)}/{y}, 0; x)
\nn
&
            - G(-((y-1) y), -{(1-y)^{2}}/{y}; x)
            - G(1-y, 0; x)
            - G(1-y, 1; x)
            + G(0, 0; x)
            + G(0, 1; x)
            - G(1, 0; x)
\nn
&
            - G(1, 1; x)
            - G(0, 0; y)
            - 2\, G(0, 1; y)
            - 2\, G(1, 0; y)
            \,\big]
            + \mathcal{O}(\epsilon^{3}) \,,
\nn
g_{8} = {} &
            \mathcal{O}(\epsilon^{3}) \,,
\nn
g_{9} = {} &
            \mathcal{O}(\epsilon^{3}) \,,
\nn
g_{10} = {} &
            2\, \epsilon^{2}\,
            \big[\,
            G(0; x)\, G(0; y)
            + G(1; x)\, G(0; y)
            - 2\, G(0; x)\, G(1; y)
            - 2\, G(1; x)\, G(1; y)
            - G(0; y)\, G(-{(1-y)^{2}}/{y}; x)
\nn
&
            + 2\, G(1; y)\, G(-{(1-y)^{2}}/{y}; x)
            - G(0, -{(1-y)^{2}}/{y}; x)
            - G(1, -{(1-y)^{2}}/{y}; x)
            - G(-{(1-y)^{2}}/{y}, 0; x)
\nn
&
            + G(-{(1-y)^{2}}/{y}, -{(1-y)^{2}}/{y}; x)
            - G(-{(1-y)^{2}}/{y}, 1; x)
            - 2\, G(0, 1; y)
            - 2\, G(1, 0; y)
            + 4\, G(1, 1; y)
\nn
&
            + G(0, 0; x)
            + G(0, 1; x)
            + G(1, 0; x)
            + G(1, 1; x)
            + G(0, 0; y)
            \,\big]
            + \mathcal{O}(\epsilon^{3}) \,,
\nn
g_{11} = {} &
            \epsilon\,
            \big[\,
            2\, G(1; y)
            - G(0; y)
            - G(0; x)
            - G(1; x)
            + G(-{(1-y)^{2}}/{y}; x)
            \,\big]
\nn
&
            + \epsilon^{2}\,
            \big[\,
            8\, G(0; x)\, G(1; y)
            + 8\, G(1; x)\, G(1; y)
            - 4\, G(0; y)\, G(-{(1-y)^{2}}/{y}; x)
            + 8\, G(1; y)\, G(-{(1-y)^{2}}/{y}; x)
\nn
&
            + 6\, G(0; y)\, G(1+\sqrt{y+{1}/{y}-1}; x)
            + 2\, G(0; y)\, G(\sqrt{-{(1-y)^{2}}/{y}}; x)
            + 2\, G(0; y)\, G(-\sqrt{-{(1-y)^{2}}/{y}}; x)
\nn
&
            + 6\, G(0; y)\, G(1-\sqrt{y+{1}/{y}-1}; x)
            - 4\, G(1; y)\, G(\sqrt{-{(1-y)^{2}}/{y}}; x)
            - 4\, G(1; y)\, G(-\sqrt{-{(1-y)^{2}}/{y}}; x)
\nn
&
            - 12\, G(1; y)\, G(1+\sqrt{y+{1}/{y}-1}; x)
            + 6\, G(1+\sqrt{y+{1}/{y}-1}, 0; x)
            + 6\, G(1-\sqrt{y+{1}/{y}-1}, 1; x)
\nn
&
            - 12\, G(1; y)\, G(1-\sqrt{y+{1}/{y}-1}; x)
            + 6\, G(1-\sqrt{y+{1}/{y}-1}, 0; x)
            + 6\, G(1-\sqrt{y+{1}/{y}-1}, 1; x)
\nn
&
            - 4\, G(0; x)\, G(0; y)
            - 2\, G(-\sqrt{-{(1-y)^{2}}/{y}}, -{(1-y)^{2}}/{y}; x)
            - 2\, G(\sqrt{-{(1-y)^{2}}/{y}}, -{(1-y)^{2}}/{y}; x)
\nn
&
            - 4\, G(1; x)\, G(0; y)
            - 6\, G(1-\sqrt{y+{1}/{y}-1}, -{(1-y)^{2}}/{y}; x)
            - 6\, G(1+\sqrt{y+{1}/{y}-1}, -{(1-y)^{2}}/{y}; x)
\nn
&
            + 2\, G(\sqrt{-{(1-y)^{2}}/{y}}, 0; x)
            + 2\, G(-\sqrt{-{(1-y)^{2}}/{y}}, 0; x)
            + 4\, G(0, -{(1-y)^{2}}/{y}; x)
            + 4\, G(1, -{(1-y)^{2}}/{y}; x)
\nn
&
            + 2\, G(\sqrt{-{(1-y)^{2}}/{y}}, 1; x)
            + 2\, G(-\sqrt{-{(1-y)^{2}}/{y}}, 1; x)
            - 4\, G(-{(1-y)^{2}}/{y}, 0; x)
            - 4\, G(-{(1-y)^{2}}/{y}, 1; x)
\nn
&
            + 4\, G(-{(1-y)^{2}}/{y}, -{(1-y)^{2}}/{y}; x)
            - 4\, G(0, 0; x)
            - 4\, G(0, 1; x)
            - 4\, G(1, 0; x)
            - 4\, G(1, 1; x)
            - 4\, G(0, 0; y)
\nn
&
            + 8\, G(0, 1; y)
            + 8\, G(1, 0; y)
            - 16\, G(1, 1; y)
            + {\pi ^{2}}/{6}
            \,\big]
            + \mathcal{O}(\epsilon^{3}) \,,
\nn
g_{12} = {} &
            2\, \epsilon^{2}\, G(0, 0; y)
            + \mathcal{O}(\epsilon^{3}) \,,
\nn
g_{13} = {} &
            - \epsilon\, \, G(0; y)
            + \epsilon^{2}
            \big[\,
            2\, G(1, 0; y)
            - 4\, G(0, 0; y)
            + 6\, G(-1, 0; y)
            + {\pi ^{2}}/{6}
            \,\big]
            + \mathcal{O}(\epsilon^{3}) \,,
\nn
g_{14} = {} &
            \mathcal{O}(\epsilon^{3}) \,,
\nn
g_{15} = {} &
            {1}/{2}\, \epsilon^{2}\,
            \big[\,
            G(1; x)\, G(0; y)
            - G(0; x)\, G(0; y)
            + 2\, G(0; x)\, G(1; y)
            - 2\, G(1; x)\, G(1; y)
            - 2\, G(1; y)\, G(1-y; x)
\nn
&
            - G(0; y)\, G(1-y; x)
            - G({(y-1)}/{y^{2}}, 0; x)
            - 3\, G(0; y)\, G({(y-1)}/{y^{2}}; x)
            + 2\, G(1; y)\, G({(y-1)}/{y^{2}}; x)
\nn
&
            + 3\, G(0; y)\, G({(y-1)}/{y}; x)
            + G(0; y)\, G(-{(1-y)^{2}}/{y}; x)
            + G(0; y)\, G(-((y-1) y); x)
            + G(1-y, 0; x)
\nn
&
            - 2\, G(1; y)\, G({(y-1)}/{y}; x)
            + 2\, G(1; y)\, G(-((y-1) y); x)
            - 2\, G(1; y)\, G(-{(1-y)^{2}}/{y}; x)
            + 2\, G(1, 0; y)
\nn
&
            + G({(y-1)}/{y^{2}}, -{(1-y)^{2}}/{y}; x)
            + G(0, -{(1-y)^{2}}/{y}; x)
            - G(1, -{(1-y)^{2}}/{y}; x)
            - G({(y-1)}/{y^{2}}, 1; x)
\nn
&
            - G(-{(1-y)^{2}}/{y}, -{(1-y)^{2}}/{y}; x)
            - G(-((y-1) y), 0; x)
            - G(-((y-1) y), 1; x)
            + G({(y-1)}/{y}, 1; x)
\nn
&
            + G(-((y-1) y), -{(1-y)^{2}}/{y}; x)
            + G(-{(1-y)^{2}}/{y}, 1; x)
            + G(-{(1-y)^{2}}/{y}, 0; x)
            + G({(y-1)}/{y}, 0; x)
\nn
&
            - G({(y-1)}/{y}, -{(1-y)^{2}}/{y}; x)
            - G(1-y, -{(1-y)^{2}}/{y}; x)
            + G(1-y, 1; x)
            + 2\, G(0, 1; y)
            + G(1, 1; x)
\nn
&
            - G(0, 0; x)
            - G(0, 1; x)
            + G(1, 0; x)
            + G(0, 0; y)
            - 4\, G(1, 1; y)
            \,\big]
            + \mathcal{O}(\epsilon^{3}) \,,
\nn
g_{16} = {} &
            {1}/{6}\, \epsilon^{2}\,
            \big[\,
            3\, G(0; x)\, G(0; y)
            + 3\, G(1; x)\, G(0; y)
            - 6\, G(0; x)\, G(1; y)
            - 6\, G(1; x)\, G(1; y)
            - 3\, G({(y-1)}/{y^{2}}, 1; x)
\nn
&
            + 3\, G(0; y)\, G(1-y; x)
            + 6\, G(1; y)\, G(1-y; x)
            + 6\, G(1; y)\, G({(y-1)}/{y}; x)
            + 3\, G(0; y)\, G(-((y-1) y); x)
\nn
&
            - 9\, G(0; y)\, G({(y-1)}/{y^{2}}; x)
            - 9\, G(0; y)\, G({(y-1)}/{y}; x)
            + 6\, G(1; y)\, G(-((y-1) y); x)
            - 3\, G(1-y, 0; x)
\nn
&
            + 3\, G(0; y)\, G(-{(1-y)^{2}}/{y}; x)
            - 6\, G(1; y)\, G(-{(1-y)^{2}}/{y}; x)
            + 6\, G(1; y)\, G({(y-1)}/{y^{2}}; x)
            + 15\, G(0, 0; y)
\nn
&
            + 3\, G({(y-1)}/{y}, -{(1-y)^{2}}/{y}; x)
            + 3\, G({(y-1)}/{y^{2}}, -{(1-y)^{2}}/{y}; x)
            - 3\, G(-{(1-y)^{2}}/{y}, -{(1-y)^{2}}/{y}; x)
\nn
&
            + 3\, G(-((y-1) y), -{(1-y)^{2}}/{y}; x)
            + 3\, G(1-y, -{(1-y)^{2}}/{y}; x)
            + 3\, G(-{(1-y)^{2}}/{y}, 0; x)
            - 6\, G(1, 0; y)
\nn
&
            - 3\, G(-((y-1) y), 1; x)
            - 3\, G(0, -{(1-y)^{2}}/{y}; x)
            - 3\, G(1, -{(1-y)^{2}}/{y}; x)
            + 3\, G(-{(1-y)^{2}}/{y}, 1; x)
\nn
&
            - 3\, G({(y-1)}/{y}, 0; x)
            - 3\, G({(y-1)}/{y}, 1; x)
            - 3\, G({(y-1)}/{y^{2}}, 0; x)
            - 3\, G(1-y, 1; x)
            + 12\, G(1, 1; y)
\nn
&
            - 3\, G(-((y-1) y), 0; x)
            + 3\, G(0, 0; x)
            + 3\, G(0, 1; x)
            + 3\, G(1, 0; x)
            + 3\, G(1, 1; x)
            - 6\, G(0, 1; y)
            + \pi ^{2} \,\big]
\nn
&
            + \mathcal{O}(\epsilon^{3}) \,,
\nn
g_{17} = {} &
            \mathcal{O}(\epsilon^{3}) \,,
\nn
g_{18} = {} &
            {1}/{2}\, \epsilon^{2}\,
            \big[\,
            G(1; x)\, G(0; y)
            - G(0; x)\, G(0; y)
            + 2\, G(0; x)\, G(1; y)
            - 2\, G(1; x)\, G(1; y)
            - 2\, G(1; y)\, G(1-y; x)
\nn
&
            - G(0; y)\, G(1-y; x)
            - 2\, G(1; y)\, G(-{(1-y)^{2}}/{y}; x)
            + 2\, G(1; y)\, G(-((y-1) y); x)
            + G({(y-1)}/{y}, 0; x)
\nn
&
            + G(0; y)\, G(-((y-1) y); x)
            + G(0; y)\, G(-{(1-y)^{2}}/{y}; x)
            - 2\, G(1; y)\, G({(y-1)}/{y}; x)
            + G(1-y, 0; x)
\nn
&
            + 3\, G(0; y)\, G({(y-1)}/{y}; x)
            - 3\, G(0; y)\, G({(y-1)}/{y^{2}}; x)
            + 2\, G(1; y)\, G({(y-1)}/{y^{2}}; x)
            + G(1-y, 1; x)
\nn
&
            - G({(y-1)}/{y}, -{(1-y)^{2}}/{y}; x)
            - G(-{(1-y)^{2}}/{y}, -{(1-y)^{2}}/{y}; x)
            + G({(y-1)}/{y}, 1; x)
            - 4\, G(1, 1; y)
\nn
&
            + G({(y-1)}/{y^{2}}, -{(1-y)^{2}}/{y}; x)
            + G(-((y-1) y), -{(1-y)^{2}}/{y}; x)
            - G({(y-1)}/{y^{2}}, 0; x) + G(1, 1; x)
\nn
&
            - G(1-y, -{(1-y)^{2}}/{y}; x)
            - G({(y-1)}/{y^{2}}, 1; x)
            - G(-((y-1) y), 1; x)
            - G(-((y-1) y), 0; x)
\nn
&
            + G(-{(1-y)^{2}}/{y}, 0; x)
            + G(-{(1-y)^{2}}/{y}, 1; x)
            - G(1, -{(1-y)^{2}}/{y}; x)
            + G(0, -{(1-y)^{2}}/{y}; x)
\nn
&
            - G(0, 0; x)
            - G(0, 1; x)
            + G(1, 0; x)
            + G(0, 0; y)
            + 2\, G(0, 1; y)
            + 2\, G(1, 0; y)
            \,\big]
            + \mathcal{O}(\epsilon^{3}) \,,
\nn
g_{19} = {} &
            {1}/{6}\, \epsilon^{2}\,
            \big[\,
            3\, G(0; y)\, G(1-y; x)
            - 6\, G(0; x)\, G(0; y)
            + 6\, G(1; y)\, G(1-y; x)
            - 6\, G(0; y)\, G(-{(1-y)^{2}}/{y}; x)
\nn
&
            - 6\, G(1; x)\, G(0; y)
            - 6\, G(1; y)\, G({(y-1)}/{y}; x)
            + 9\, G(0; y)\, G({(y-1)}/{y}; x)
            - 6\, G(1; y)\, G({(y-1)}/{y^{2}}; x)
\nn
&
            + 3\, G(0; y)\, G(-((y-1) y); x)
            + 6\, G(1; y)\, G(-((y-1) y); x)
            + 9\, G(0; y)\, G({(y-1)}/{y^{2}}; x)
            - 9\, G(0, 0; y)
\nn
&
            - 3\, G({(y-1)}/{y}, -{(1-y)^{2}}/{y}; x)
            + 3\, G(1-y, -{(1-y)^{2}}/{y}; x)
            + 3\, G({(y-1)}/{y^{2}}, 0; x)
            - 3\, G(1-y, 1; x)
\nn
&
            + 3\, G(-((y-1) y), -{(1-y)^{2}}/{y}; x)
            - 3\, G({(y-1)}/{y^{2}}, -{(1-y)^{2}}/{y}; x)
            - 3\, G(1-y, 0; x)+12\, G(0, 1; y)
\nn
&
            - 3\, G(-((y-1) y), 0; x)
            - 3\, G(-((y-1) y), 1; x)
            + 3\, G({(y-1)}/{y^{2}}, 1; x)
            + 3\, G({(y-1)}/{y}, 0; x)
\nn
&
            + 3\, G({(y-1)}/{y}, 1; x)
            + 6\, G(1, 0; y)
            + \pi ^{2}
            \,\big]
            + \mathcal{O}(\epsilon^{3}) \,,
\nn
g_{20} = {} &
            \mathcal{O}(\epsilon^{3}) \,,
\nn
g_{21} = {} &
            \mathcal{O}(\epsilon^{3}) \,,
\nn
g_{22} = {} &
            \mathcal{O}(\epsilon^{3}) \,,
\nn
g_{23} = {} &
            - 2\, \epsilon^{2}\,
            \big[\,
            G(0; x)\, G(0; y)
            + G(1; x)\, G(0; y)
            - 2\, G(0; x)\, G(1; y)
            - 2\, G(1; x)\, G(1; y)
            - G(0, -{(1-y)^{2}}/{y}; x)
\nn
&
            - G(0; y)\, G(-{(1-y)^{2}}/{y}; x)
            + 2\, G(1; y)\, G(-{(1-y)^{2}}/{y}; x)
            + G(-{(1-y)^{2}}/{y}, -{(1-y)^{2}}/{y}; x)
\nn
&
            - G(1, -{(1-y)^{2}}/{y}; x)
            - G(-{(1-y)^{2}}/{y}, 0; x)
            - G(-{(1-y)^{2}}/{y}, 1; x)
            + G(0, 0; x)
            + G(0, 1; x)
\nn
&
            + G(1, 0; x)
            + G(1, 1; x)
            + G(0, 0; y)
            - 2\, G(0, 1; y)
            - 2\, G(1, 0; y)
            + 4\, G(1, 1; y)
            \,\big]
            + \mathcal{O}(\epsilon^{3}) \,,
\nn
g_{24} = {} &
            \mathcal{O}(\epsilon^{3}) \,,
\nn
g_{25} = {} &
            \mathcal{O}(\epsilon^{3}) \,,
\nn
g_{26} = {} &
            2\, \epsilon^{2}\,
            \big[\,
            G(0; x)\, G(0; y)
            + G(1; x)\, G(0; y)
            - 2\, G(0; x)\, G(1; y)
            - G(0, -{(1-y)^{2}}/{y}; x)
            - G(1, -{(1-y)^{2}}/{y}; x)
\nn
&
            - 2\, G(1; x)\, G(1; y)
            - G(0; y)\, G(-{(1-y)^{2}}/{y}; x)
            + 2\, G(1; y)\, G(-{(1-y)^{2}}/{y}; x)
            - G(-{(1-y)^{2}}/{y}, 0; x)
\nn
&
            + G(-{(1-y)^{2}}/{y}, -{(1-y)^{2}}/{y}; x)
            - G(-{(1-y)^{2}}/{y}, 1; x)
            - 2\, G(0, 1; y)
            - 2\, G(1, 0; y)
            + 4\, G(1,  1; y)
\nn
&
            + G(0, 0; x)
            + G(0, 1; x)
            + G(1, 0; x)
            + G(1, 1; x)
            + G(0, 0; y)
            \,\big]
            + \mathcal{O}(\epsilon^{3}) \,.
\nonumber
\end{align}
}

\bibliographystyle{apsrev4-2}
\bibliography{refs}

\begin{thebibliography}{109}%
\makeatletter
\providecommand \@ifxundefined [1]{%
 \@ifx{#1\undefined}
}%
\providecommand \@ifnum [1]{%
 \ifnum #1\expandafter \@firstoftwo
 \else \expandafter \@secondoftwo
 \fi
}%
\providecommand \@ifx [1]{%
 \ifx #1\expandafter \@firstoftwo
 \else \expandafter \@secondoftwo
 \fi
}%
\providecommand \natexlab [1]{#1}%
\providecommand \enquote  [1]{``#1''}%
\providecommand \bibnamefont  [1]{#1}%
\providecommand \bibfnamefont [1]{#1}%
\providecommand \citenamefont [1]{#1}%
\providecommand \href@noop [0]{\@secondoftwo}%
\providecommand \href [0]{\begingroup \@sanitize@url \@href}%
\providecommand \@href[1]{\@@startlink{#1}\@@href}%
\providecommand \@@href[1]{\endgroup#1\@@endlink}%
\providecommand \@sanitize@url [0]{\catcode `\\12\catcode `\$12\catcode
  `\&12\catcode `\#12\catcode `\^12\catcode `\_12\catcode `\%12\relax}%
\providecommand \@@startlink[1]{}%
\providecommand \@@endlink[0]{}%
\providecommand \url  [0]{\begingroup\@sanitize@url \@url }%
\providecommand \@url [1]{\endgroup\@href {#1}{\urlprefix }}%
\providecommand \urlprefix  [0]{URL }%
\providecommand \Eprint [0]{\href }%
\providecommand \doibase [0]{https://doi.org/}%
\providecommand \selectlanguage [0]{\@gobble}%
\providecommand \bibinfo  [0]{\@secondoftwo}%
\providecommand \bibfield  [0]{\@secondoftwo}%
\providecommand \translation [1]{[#1]}%
\providecommand \BibitemOpen [0]{}%
\providecommand \bibitemStop [0]{}%
\providecommand \bibitemNoStop [0]{.\EOS\space}%
\providecommand \EOS [0]{\spacefactor3000\relax}%
\providecommand \BibitemShut  [1]{\csname bibitem#1\endcsname}%
\let\auto@bib@innerbib\@empty
\bibitem [{\citenamefont {Aad}\ \emph {et~al.}(2012{\natexlab{a}})\citenamefont
  {Aad} \emph {et~al.}}]{ATLAS:2012yve}%
  \BibitemOpen
  \bibfield  {author} {\bibinfo {author} {\bibfnamefont {G.}~\bibnamefont
  {Aad}} \emph {et~al.} (\bibinfo {collaboration} {ATLAS}),\ }\href
  {https://doi.org/10.1016/j.physletb.2012.08.020} {\bibfield  {journal}
  {\bibinfo  {journal} {Phys. Lett. B}\ }\textbf {\bibinfo {volume} {716}},\
  \bibinfo {pages} {1} (\bibinfo {year} {2012}{\natexlab{a}})},\ \Eprint
  {https://arxiv.org/abs/1207.7214} {arXiv:1207.7214 [hep-ex]} \BibitemShut
  {NoStop}%
\bibitem [{\citenamefont {Chatrchyan}\ \emph {et~al.}(2012)\citenamefont
  {Chatrchyan} \emph {et~al.}}]{CMS:2012qbp}%
  \BibitemOpen
  \bibfield  {author} {\bibinfo {author} {\bibfnamefont {S.}~\bibnamefont
  {Chatrchyan}} \emph {et~al.} (\bibinfo {collaboration} {CMS}),\ }\href
  {https://doi.org/10.1016/j.physletb.2012.08.021} {\bibfield  {journal}
  {\bibinfo  {journal} {Phys. Lett. B}\ }\textbf {\bibinfo {volume} {716}},\
  \bibinfo {pages} {30} (\bibinfo {year} {2012})},\ \Eprint
  {https://arxiv.org/abs/1207.7235} {arXiv:1207.7235 [hep-ex]} \BibitemShut
  {NoStop}%
\bibitem [{\citenamefont {Barate}\ \emph {et~al.}(1999)\citenamefont {Barate}
  \emph {et~al.}}]{ALEPH:1999ytd}%
  \BibitemOpen
  \bibfield  {author} {\bibinfo {author} {\bibfnamefont {R.}~\bibnamefont
  {Barate}} \emph {et~al.} (\bibinfo {collaboration} {ALEPH}),\ }\href
  {https://doi.org/10.1016/S0370-2693(99)01288-5} {\bibfield  {journal}
  {\bibinfo  {journal} {Phys. Lett. B}\ }\textbf {\bibinfo {volume} {469}},\
  \bibinfo {pages} {287} (\bibinfo {year} {1999})},\ \Eprint
  {https://arxiv.org/abs/hep-ex/9911003} {arXiv:hep-ex/9911003} \BibitemShut
  {NoStop}%
\bibitem [{\citenamefont {Achard}\ \emph {et~al.}(2003)\citenamefont {Achard}
  \emph {et~al.}}]{L3:2003kow}%
  \BibitemOpen
  \bibfield  {author} {\bibinfo {author} {\bibfnamefont {P.}~\bibnamefont
  {Achard}} \emph {et~al.} (\bibinfo {collaboration} {L3}),\ }\href
  {https://doi.org/10.1016/j.physletb.2003.08.023} {\bibfield  {journal}
  {\bibinfo  {journal} {Phys. Lett. B}\ }\textbf {\bibinfo {volume} {572}},\
  \bibinfo {pages} {133} (\bibinfo {year} {2003})},\ \Eprint
  {https://arxiv.org/abs/hep-ex/0308013} {arXiv:hep-ex/0308013} \BibitemShut
  {NoStop}%
\bibitem [{\citenamefont {Abdallah}\ \emph {et~al.}(2003)\citenamefont
  {Abdallah} \emph {et~al.}}]{DELPHI:2003kgg}%
  \BibitemOpen
  \bibfield  {author} {\bibinfo {author} {\bibfnamefont {J.}~\bibnamefont
  {Abdallah}} \emph {et~al.} (\bibinfo {collaboration} {DELPHI}),\ }\href
  {https://doi.org/10.1140/epjc/s2003-01287-0} {\bibfield  {journal} {\bibinfo
  {journal} {Eur. Phys. J. C}\ }\textbf {\bibinfo {volume} {30}},\ \bibinfo
  {pages} {447} (\bibinfo {year} {2003})},\ \Eprint
  {https://arxiv.org/abs/hep-ex/0307050} {arXiv:hep-ex/0307050} \BibitemShut
  {NoStop}%
\bibitem [{\citenamefont {Alcaraz}\ \emph {et~al.}(2006)\citenamefont {Alcaraz}
  \emph {et~al.}}]{ALEPH:2006bhb}%
  \BibitemOpen
  \bibfield  {author} {\bibinfo {author} {\bibfnamefont {J.}~\bibnamefont
  {Alcaraz}} \emph {et~al.} (\bibinfo {collaboration} {ALEPH, DELPHI, L3, OPAL,
  LEP Electroweak Working Group}),\ }\href@noop {} {\  (\bibinfo {year}
  {2006})},\ \Eprint {https://arxiv.org/abs/hep-ex/0612034}
  {arXiv:hep-ex/0612034} \BibitemShut {NoStop}%
\bibitem [{\citenamefont {Abdallah}\ \emph {et~al.}(2007)\citenamefont
  {Abdallah} \emph {et~al.}}]{DELPHI:2007gzg}%
  \BibitemOpen
  \bibfield  {author} {\bibinfo {author} {\bibfnamefont {J.}~\bibnamefont
  {Abdallah}} \emph {et~al.} (\bibinfo {collaboration} {DELPHI}),\ }\href
  {https://doi.org/10.1140/epjc/s10052-007-0345-0} {\bibfield  {journal}
  {\bibinfo  {journal} {Eur. Phys. J. C}\ }\textbf {\bibinfo {volume} {51}},\
  \bibinfo {pages} {525} (\bibinfo {year} {2007})},\ \Eprint
  {https://arxiv.org/abs/0706.2741} {arXiv:0706.2741 [hep-ex]} \BibitemShut
  {NoStop}%
\bibitem [{\citenamefont {Schael}\ \emph {et~al.}(2009)\citenamefont {Schael}
  \emph {et~al.}}]{ALEPH:2009rug}%
  \BibitemOpen
  \bibfield  {author} {\bibinfo {author} {\bibfnamefont {S.}~\bibnamefont
  {Schael}} \emph {et~al.} (\bibinfo {collaboration} {ALEPH}),\ }\href
  {https://doi.org/10.1088/1126-6708/2009/04/124} {\bibfield  {journal}
  {\bibinfo  {journal} {JHEP}\ }\textbf {\bibinfo {volume} {04}},\ \bibinfo
  {pages} {124}}\BibitemShut {NoStop}%
\bibitem [{\citenamefont {Aaltonen}\ \emph {et~al.}(2008)\citenamefont
  {Aaltonen} \emph {et~al.}}]{CDF:2008eiu}%
  \BibitemOpen
  \bibfield  {author} {\bibinfo {author} {\bibfnamefont {T.~A.}\ \bibnamefont
  {Aaltonen}} \emph {et~al.} (\bibinfo {collaboration} {CDF}),\ }\href
  {https://doi.org/10.1103/PhysRevLett.100.201801} {\bibfield  {journal}
  {\bibinfo  {journal} {Phys. Rev. Lett.}\ }\textbf {\bibinfo {volume} {100}},\
  \bibinfo {pages} {201801} (\bibinfo {year} {2008})},\ \Eprint
  {https://arxiv.org/abs/0801.4806} {arXiv:0801.4806 [hep-ex]} \BibitemShut
  {NoStop}%
\bibitem [{\citenamefont {Abazov}\ \emph {et~al.}(2008)\citenamefont {Abazov}
  \emph {et~al.}}]{D0:2008sin}%
  \BibitemOpen
  \bibfield  {author} {\bibinfo {author} {\bibfnamefont {V.~M.}\ \bibnamefont
  {Abazov}} \emph {et~al.} (\bibinfo {collaboration} {D0}),\ }\href
  {https://doi.org/10.1103/PhysRevLett.101.171803} {\bibfield  {journal}
  {\bibinfo  {journal} {Phys. Rev. Lett.}\ }\textbf {\bibinfo {volume} {101}},\
  \bibinfo {pages} {171803} (\bibinfo {year} {2008})},\ \Eprint
  {https://arxiv.org/abs/0808.0703} {arXiv:0808.0703 [hep-ex]} \BibitemShut
  {NoStop}%
\bibitem [{\citenamefont {Aaltonen}\ \emph {et~al.}(2014)\citenamefont
  {Aaltonen} \emph {et~al.}}]{CDF:2014nef}%
  \BibitemOpen
  \bibfield  {author} {\bibinfo {author} {\bibfnamefont {T.~A.}\ \bibnamefont
  {Aaltonen}} \emph {et~al.} (\bibinfo {collaboration} {CDF}),\ }\href
  {https://doi.org/10.1103/PhysRevD.89.112001} {\bibfield  {journal} {\bibinfo
  {journal} {Phys. Rev. D}\ }\textbf {\bibinfo {volume} {89}},\ \bibinfo
  {pages} {112001} (\bibinfo {year} {2014})},\ \Eprint
  {https://arxiv.org/abs/1403.2300} {arXiv:1403.2300 [hep-ex]} \BibitemShut
  {NoStop}%
\bibitem [{\citenamefont {Tumasyan}\ \emph {et~al.}(2021)\citenamefont
  {Tumasyan} \emph {et~al.}}]{CMS:2021pqj}%
  \BibitemOpen
  \bibfield  {author} {\bibinfo {author} {\bibfnamefont {A.}~\bibnamefont
  {Tumasyan}} \emph {et~al.} (\bibinfo {collaboration} {CMS}),\ }\href
  {https://doi.org/10.1103/PhysRevLett.127.191801} {\bibfield  {journal}
  {\bibinfo  {journal} {Phys. Rev. Lett.}\ }\textbf {\bibinfo {volume} {127}},\
  \bibinfo {pages} {191801} (\bibinfo {year} {2021})},\ \Eprint
  {https://arxiv.org/abs/2107.01137} {arXiv:2107.01137 [hep-ex]} \BibitemShut
  {NoStop}%
\bibitem [{\citenamefont {Aad}\ \emph {et~al.}(2012{\natexlab{b}})\citenamefont
  {Aad} \emph {et~al.}}]{ATLAS:2011ahl}%
  \BibitemOpen
  \bibfield  {author} {\bibinfo {author} {\bibfnamefont {G.}~\bibnamefont
  {Aad}} \emph {et~al.} (\bibinfo {collaboration} {ATLAS}),\ }\href
  {https://doi.org/10.1103/PhysRevLett.108.041804} {\bibfield  {journal}
  {\bibinfo  {journal} {Phys. Rev. Lett.}\ }\textbf {\bibinfo {volume} {108}},\
  \bibinfo {pages} {041804} (\bibinfo {year} {2012}{\natexlab{b}})},\ \Eprint
  {https://arxiv.org/abs/1110.5016} {arXiv:1110.5016 [hep-ex]} \BibitemShut
  {NoStop}%
\bibitem [{\citenamefont {Chatrchyan}\ \emph {et~al.}(2013)\citenamefont
  {Chatrchyan} \emph {et~al.}}]{CMS:2013piy}%
  \BibitemOpen
  \bibfield  {author} {\bibinfo {author} {\bibfnamefont {S.}~\bibnamefont
  {Chatrchyan}} \emph {et~al.} (\bibinfo {collaboration} {CMS}),\ }\href
  {https://doi.org/10.1016/j.physletb.2013.03.027} {\bibfield  {journal}
  {\bibinfo  {journal} {Phys. Lett. B}\ }\textbf {\bibinfo {volume} {721}},\
  \bibinfo {pages} {190} (\bibinfo {year} {2013})},\ \Eprint
  {https://arxiv.org/abs/1301.4698} {arXiv:1301.4698 [hep-ex]} \BibitemShut
  {NoStop}%
\bibitem [{\citenamefont {Khachatryan}\ \emph {et~al.}(2015)\citenamefont
  {Khachatryan} \emph {et~al.}}]{CMS:2015qgb}%
  \BibitemOpen
  \bibfield  {author} {\bibinfo {author} {\bibfnamefont {V.}~\bibnamefont
  {Khachatryan}} \emph {et~al.} (\bibinfo {collaboration} {CMS}),\ }\href
  {https://doi.org/10.1140/epjc/s10052-015-3706-0} {\bibfield  {journal}
  {\bibinfo  {journal} {Eur. Phys. J. C}\ }\textbf {\bibinfo {volume} {75}},\
  \bibinfo {pages} {511} (\bibinfo {year} {2015})},\ \Eprint
  {https://arxiv.org/abs/1503.05467} {arXiv:1503.05467 [hep-ex]} \BibitemShut
  {NoStop}%
\bibitem [{\citenamefont {Aaboud}\ \emph {et~al.}(2017)\citenamefont {Aaboud}
  \emph {et~al.}}]{ATLAS:2016bxw}%
  \BibitemOpen
  \bibfield  {author} {\bibinfo {author} {\bibfnamefont {M.}~\bibnamefont
  {Aaboud}} \emph {et~al.} (\bibinfo {collaboration} {ATLAS}),\ }\href
  {https://doi.org/10.1007/JHEP01(2017)099} {\bibfield  {journal} {\bibinfo
  {journal} {JHEP}\ }\textbf {\bibinfo {volume} {01}},\ \bibinfo {pages}
  {099}},\ \Eprint {https://arxiv.org/abs/1610.07585} {arXiv:1610.07585
  [hep-ex]} \BibitemShut {NoStop}%
\bibitem [{\citenamefont {Aad}\ \emph {et~al.}(2016)\citenamefont {Aad} \emph
  {et~al.}}]{ATLAS:2015guq}%
  \BibitemOpen
  \bibfield  {author} {\bibinfo {author} {\bibfnamefont {G.}~\bibnamefont
  {Aad}} \emph {et~al.} (\bibinfo {collaboration} {ATLAS}),\ }\href
  {https://doi.org/10.1103/PhysRevLett.116.101801} {\bibfield  {journal}
  {\bibinfo  {journal} {Phys. Rev. Lett.}\ }\textbf {\bibinfo {volume} {116}},\
  \bibinfo {pages} {101801} (\bibinfo {year} {2016})},\ \Eprint
  {https://arxiv.org/abs/1512.05314} {arXiv:1512.05314 [hep-ex]} \BibitemShut
  {NoStop}%
\bibitem [{\citenamefont {Khachatryan}\ \emph {et~al.}(2016)\citenamefont
  {Khachatryan} \emph {et~al.}}]{CMS:2016ogx}%
  \BibitemOpen
  \bibfield  {author} {\bibinfo {author} {\bibfnamefont {V.}~\bibnamefont
  {Khachatryan}} \emph {et~al.} (\bibinfo {collaboration} {CMS}),\ }\href
  {https://doi.org/10.1016/j.physletb.2016.10.054} {\bibfield  {journal}
  {\bibinfo  {journal} {Phys. Lett. B}\ }\textbf {\bibinfo {volume} {763}},\
  \bibinfo {pages} {280} (\bibinfo {year} {2016})},\ \bibinfo {note} {[Erratum:
  Phys.Lett.B 772, 884-884 (2017)]},\ \Eprint
  {https://arxiv.org/abs/1607.08834} {arXiv:1607.08834 [hep-ex]} \BibitemShut
  {NoStop}%
\bibitem [{\citenamefont {Sirunyan}\ \emph {et~al.}(2021)\citenamefont
  {Sirunyan} \emph {et~al.}}]{CMS:2020gtj}%
  \BibitemOpen
  \bibfield  {author} {\bibinfo {author} {\bibfnamefont {A.~M.}\ \bibnamefont
  {Sirunyan}} \emph {et~al.} (\bibinfo {collaboration} {CMS}),\ }\href
  {https://doi.org/10.1140/epjc/s10052-020-08817-8} {\bibfield  {journal}
  {\bibinfo  {journal} {Eur. Phys. J. C}\ }\textbf {\bibinfo {volume} {81}},\
  \bibinfo {pages} {200} (\bibinfo {year} {2021})},\ \Eprint
  {https://arxiv.org/abs/2009.01186} {arXiv:2009.01186 [hep-ex]} \BibitemShut
  {NoStop}%
\bibitem [{\citenamefont {Behnke}\ \emph {et~al.}(2013)\citenamefont {Behnke},
  \citenamefont {Brau}, \citenamefont {Foster}, \citenamefont {Fuster},
  \citenamefont {Harrison}, \citenamefont {Paterson}, \citenamefont {Peskin},
  \citenamefont {Stanitzki}, \citenamefont {Walker},\ and\ \citenamefont
  {Yamamoto}}]{Behnke:2013xla}%
  \BibitemOpen
  \bibfield  {author} {\bibinfo {author} {\bibfnamefont {T.}~\bibnamefont
  {Behnke}}, \bibinfo {author} {\bibfnamefont {J.~E.}\ \bibnamefont {Brau}},
  \bibinfo {author} {\bibfnamefont {B.}~\bibnamefont {Foster}}, \bibinfo
  {author} {\bibfnamefont {J.}~\bibnamefont {Fuster}}, \bibinfo {author}
  {\bibfnamefont {M.}~\bibnamefont {Harrison}}, \bibinfo {author}
  {\bibfnamefont {J.~M.}\ \bibnamefont {Paterson}}, \bibinfo {author}
  {\bibfnamefont {M.}~\bibnamefont {Peskin}}, \bibinfo {author} {\bibfnamefont
  {M.}~\bibnamefont {Stanitzki}}, \bibinfo {author} {\bibfnamefont
  {N.}~\bibnamefont {Walker}},\ and\ \bibinfo {author} {\bibfnamefont
  {H.}~\bibnamefont {Yamamoto}} (\bibinfo {collaboration} {ILC}),\ }\href@noop
  {} {\  (\bibinfo {year} {2013})},\ \Eprint {https://arxiv.org/abs/1306.6327}
  {arXiv:1306.6327 [physics.acc-ph]} \BibitemShut {NoStop}%
\bibitem [{\citenamefont {Baer}\ \emph {et~al.}(2013)\citenamefont {Baer} \emph
  {et~al.}}]{ILC:2013jhg}%
  \BibitemOpen
  \bibfield  {author} {\bibinfo {author} {\bibfnamefont {H.}~\bibnamefont
  {Baer}} \emph {et~al.} (\bibinfo {collaboration} {ILC}),\ }\href@noop {} {\
  (\bibinfo {year} {2013})},\ \Eprint {https://arxiv.org/abs/1306.6352}
  {arXiv:1306.6352 [hep-ph]} \BibitemShut {NoStop}%
\bibitem [{\citenamefont {Bambade}\ \emph {et~al.}(2019)\citenamefont {Bambade}
  \emph {et~al.}}]{Bambade:2019fyw}%
  \BibitemOpen
  \bibfield  {author} {\bibinfo {author} {\bibfnamefont {P.}~\bibnamefont
  {Bambade}} \emph {et~al.},\ }\href@noop {} {\  (\bibinfo {year} {2019})},\
  \Eprint {https://arxiv.org/abs/1903.01629} {arXiv:1903.01629 [hep-ex]}
  \BibitemShut {NoStop}%
\bibitem [{CEP(2018)}]{CEPCStudyGroup:2018rmc}%
  \BibitemOpen
  \href@noop {} {\  (\bibinfo {year} {2018})},\ \Eprint
  {https://arxiv.org/abs/1809.00285} {arXiv:1809.00285 [physics.acc-ph]}
  \BibitemShut {NoStop}%
\bibitem [{\citenamefont {Dong}\ \emph {et~al.}(2018)\citenamefont {Dong} \emph
  {et~al.}}]{CEPCStudyGroup:2018ghi}%
  \BibitemOpen
  \bibfield  {author} {\bibinfo {author} {\bibfnamefont {M.}~\bibnamefont
  {Dong}} \emph {et~al.} (\bibinfo {collaboration} {CEPC Study Group}),\
  }\href@noop {} {\  (\bibinfo {year} {2018})},\ \Eprint
  {https://arxiv.org/abs/1811.10545} {arXiv:1811.10545 [hep-ex]} \BibitemShut
  {NoStop}%
\bibitem [{\citenamefont {Abada}\ \emph
  {et~al.}(2019{\natexlab{a}})\citenamefont {Abada} \emph
  {et~al.}}]{FCC:2018byv}%
  \BibitemOpen
  \bibfield  {author} {\bibinfo {author} {\bibfnamefont {A.}~\bibnamefont
  {Abada}} \emph {et~al.} (\bibinfo {collaboration} {FCC}),\ }\href
  {https://doi.org/10.1140/epjc/s10052-019-6904-3} {\bibfield  {journal}
  {\bibinfo  {journal} {Eur. Phys. J. C}\ }\textbf {\bibinfo {volume} {79}},\
  \bibinfo {pages} {474} (\bibinfo {year} {2019}{\natexlab{a}})}\BibitemShut
  {NoStop}%
\bibitem [{\citenamefont {Abada}\ \emph
  {et~al.}(2019{\natexlab{b}})\citenamefont {Abada} \emph
  {et~al.}}]{FCC:2018evy}%
  \BibitemOpen
  \bibfield  {author} {\bibinfo {author} {\bibfnamefont {A.}~\bibnamefont
  {Abada}} \emph {et~al.} (\bibinfo {collaboration} {FCC}),\ }\href
  {https://doi.org/10.1140/epjst/e2019-900045-4} {\bibfield  {journal}
  {\bibinfo  {journal} {Eur. Phys. J. Special Topic}\ }\textbf {\bibinfo
  {volume} {228}},\ \bibinfo {pages} {261} (\bibinfo {year}
  {2019}{\natexlab{b}})}\BibitemShut {NoStop}%
\bibitem [{\citenamefont {Brown}\ and\ \citenamefont
  {Mikaelian}(1979)}]{Brown:1978mq}%
  \BibitemOpen
  \bibfield  {author} {\bibinfo {author} {\bibfnamefont {R.~W.}\ \bibnamefont
  {Brown}}\ and\ \bibinfo {author} {\bibfnamefont {K.~O.}\ \bibnamefont
  {Mikaelian}},\ }\href {https://doi.org/10.1103/PhysRevD.19.922} {\bibfield
  {journal} {\bibinfo  {journal} {Phys. Rev. D}\ }\textbf {\bibinfo {volume}
  {19}},\ \bibinfo {pages} {922} (\bibinfo {year} {1979})}\BibitemShut
  {NoStop}%
\bibitem [{\citenamefont {Gaemers}\ and\ \citenamefont
  {Gounaris}(1979)}]{Gaemers:1978hg}%
  \BibitemOpen
  \bibfield  {author} {\bibinfo {author} {\bibfnamefont {K.~J.~F.}\
  \bibnamefont {Gaemers}}\ and\ \bibinfo {author} {\bibfnamefont {G.~J.}\
  \bibnamefont {Gounaris}},\ }\href {https://doi.org/10.1007/BF01440226}
  {\bibfield  {journal} {\bibinfo  {journal} {Z. Phys. C}\ }\textbf {\bibinfo
  {volume} {1}},\ \bibinfo {pages} {259} (\bibinfo {year} {1979})}\BibitemShut
  {NoStop}%
\bibitem [{\citenamefont {Denner}\ and\ \citenamefont
  {Sack}(1988)}]{Denner:1988tv}%
  \BibitemOpen
  \bibfield  {author} {\bibinfo {author} {\bibfnamefont {A.}~\bibnamefont
  {Denner}}\ and\ \bibinfo {author} {\bibfnamefont {T.}~\bibnamefont {Sack}},\
  }\href {https://doi.org/10.1016/0550-3213(88)90691-8} {\bibfield  {journal}
  {\bibinfo  {journal} {Nucl. Phys. B}\ }\textbf {\bibinfo {volume} {306}},\
  \bibinfo {pages} {221} (\bibinfo {year} {1988})}\BibitemShut {NoStop}%
\bibitem [{\citenamefont {Denner}\ and\ \citenamefont
  {Sack}(1990)}]{Denner:1988nq}%
  \BibitemOpen
  \bibfield  {author} {\bibinfo {author} {\bibfnamefont {A.}~\bibnamefont
  {Denner}}\ and\ \bibinfo {author} {\bibfnamefont {T.}~\bibnamefont {Sack}},\
  }\href {https://doi.org/10.1007/BF01549672} {\bibfield  {journal} {\bibinfo
  {journal} {Z. Phys. C}\ }\textbf {\bibinfo {volume} {45}},\ \bibinfo {pages}
  {439} (\bibinfo {year} {1990})}\BibitemShut {NoStop}%
\bibitem [{\citenamefont {Gounaris}\ \emph {et~al.}(2002)\citenamefont
  {Gounaris}, \citenamefont {Layssac},\ and\ \citenamefont
  {Renard}}]{Gounaris:2002fa}%
  \BibitemOpen
  \bibfield  {author} {\bibinfo {author} {\bibfnamefont {G.~J.}\ \bibnamefont
  {Gounaris}}, \bibinfo {author} {\bibfnamefont {J.}~\bibnamefont {Layssac}},\
  and\ \bibinfo {author} {\bibfnamefont {F.~M.}\ \bibnamefont {Renard}},\
  }\href@noop {} {\  (\bibinfo {year} {2002})},\ \Eprint
  {https://arxiv.org/abs/hep-ph/0207273} {arXiv:hep-ph/0207273} \BibitemShut
  {NoStop}%
\bibitem [{\citenamefont {Gounaris}\ \emph {et~al.}(2003)\citenamefont
  {Gounaris}, \citenamefont {Layssac},\ and\ \citenamefont
  {Renard}}]{Gounaris:2002za}%
  \BibitemOpen
  \bibfield  {author} {\bibinfo {author} {\bibfnamefont {G.~J.}\ \bibnamefont
  {Gounaris}}, \bibinfo {author} {\bibfnamefont {J.}~\bibnamefont {Layssac}},\
  and\ \bibinfo {author} {\bibfnamefont {F.~M.}\ \bibnamefont {Renard}},\
  }\href {https://doi.org/10.1103/PhysRevD.67.013012} {\bibfield  {journal}
  {\bibinfo  {journal} {Phys. Rev. D}\ }\textbf {\bibinfo {volume} {67}},\
  \bibinfo {pages} {013012} (\bibinfo {year} {2003})},\ \Eprint
  {https://arxiv.org/abs/hep-ph/0211327} {arXiv:hep-ph/0211327} \BibitemShut
  {NoStop}%
\bibitem [{\citenamefont {Demirci}\ and\ \citenamefont
  {Balantekin}(2022)}]{Demirci:2022lmr}%
  \BibitemOpen
  \bibfield  {author} {\bibinfo {author} {\bibfnamefont {M.}~\bibnamefont
  {Demirci}}\ and\ \bibinfo {author} {\bibfnamefont {A.~B.}\ \bibnamefont
  {Balantekin}},\ }\href {https://doi.org/10.1103/PhysRevD.106.073003}
  {\bibfield  {journal} {\bibinfo  {journal} {Phys. Rev. D}\ }\textbf {\bibinfo
  {volume} {106}},\ \bibinfo {pages} {073003} (\bibinfo {year} {2022})},\
  \Eprint {https://arxiv.org/abs/2209.13720} {arXiv:2209.13720 [hep-ph]}
  \BibitemShut {NoStop}%
\bibitem [{\citenamefont {Bondarenko}\ \emph {et~al.}(2024)\citenamefont
  {Bondarenko}, \citenamefont {Dydyshka}, \citenamefont {Kalinovskaya},
  \citenamefont {Sadykov},\ and\ \citenamefont
  {Yermolchyk}}]{Bondarenko:2024txj}%
  \BibitemOpen
  \bibfield  {author} {\bibinfo {author} {\bibfnamefont {S.}~\bibnamefont
  {Bondarenko}}, \bibinfo {author} {\bibfnamefont {Y.}~\bibnamefont
  {Dydyshka}}, \bibinfo {author} {\bibfnamefont {L.}~\bibnamefont
  {Kalinovskaya}}, \bibinfo {author} {\bibfnamefont {R.}~\bibnamefont
  {Sadykov}},\ and\ \bibinfo {author} {\bibfnamefont {V.}~\bibnamefont
  {Yermolchyk}},\ }\href {https://doi.org/10.1103/PhysRevD.109.033012}
  {\bibfield  {journal} {\bibinfo  {journal} {Phys. Rev. D}\ }\textbf {\bibinfo
  {volume} {109}},\ \bibinfo {pages} {033012} (\bibinfo {year} {2024})},\
  \Eprint {https://arxiv.org/abs/2401.13402} {arXiv:2401.13402 [hep-ph]}
  \BibitemShut {NoStop}%
\bibitem [{\citenamefont {Tkachov}(1981)}]{Tkachov:1981wb}%
  \BibitemOpen
  \bibfield  {author} {\bibinfo {author} {\bibfnamefont {F.~V.}\ \bibnamefont
  {Tkachov}},\ }\href {https://doi.org/10.1016/0370-2693(81)90288-4} {\bibfield
   {journal} {\bibinfo  {journal} {Phys. Lett. B}\ }\textbf {\bibinfo {volume}
  {100}},\ \bibinfo {pages} {65} (\bibinfo {year} {1981})}\BibitemShut
  {NoStop}%
\bibitem [{\citenamefont {Chetyrkin}\ and\ \citenamefont
  {Tkachov}(1981)}]{Chetyrkin:1981qh}%
  \BibitemOpen
  \bibfield  {author} {\bibinfo {author} {\bibfnamefont {K.~G.}\ \bibnamefont
  {Chetyrkin}}\ and\ \bibinfo {author} {\bibfnamefont {F.~V.}\ \bibnamefont
  {Tkachov}},\ }\href {https://doi.org/10.1016/0550-3213(81)90199-1} {\bibfield
   {journal} {\bibinfo  {journal} {Nucl. Phys. B}\ }\textbf {\bibinfo {volume}
  {192}},\ \bibinfo {pages} {159} (\bibinfo {year} {1981})}\BibitemShut
  {NoStop}%
\bibitem [{\citenamefont {Henn}(2013)}]{Henn:2013pwa}%
  \BibitemOpen
  \bibfield  {author} {\bibinfo {author} {\bibfnamefont {J.~M.}\ \bibnamefont
  {Henn}},\ }\href {https://doi.org/10.1103/PhysRevLett.110.251601} {\bibfield
  {journal} {\bibinfo  {journal} {Phys. Rev. Lett.}\ }\textbf {\bibinfo
  {volume} {110}},\ \bibinfo {pages} {251601} (\bibinfo {year} {2013})},\
  \Eprint {https://arxiv.org/abs/1304.1806} {arXiv:1304.1806 [hep-th]}
  \BibitemShut {NoStop}%
\bibitem [{\citenamefont {Henn}(2015)}]{Henn:2014qga}%
  \BibitemOpen
  \bibfield  {author} {\bibinfo {author} {\bibfnamefont {J.~M.}\ \bibnamefont
  {Henn}},\ }\href {https://doi.org/10.1088/1751-8113/48/15/153001} {\bibfield
  {journal} {\bibinfo  {journal} {J. Phys. A}\ }\textbf {\bibinfo {volume}
  {48}},\ \bibinfo {pages} {153001} (\bibinfo {year} {2015})},\ \Eprint
  {https://arxiv.org/abs/1412.2296} {arXiv:1412.2296 [hep-ph]} \BibitemShut
  {NoStop}%
\bibitem [{\citenamefont {Dlapa}\ \emph {et~al.}(2021)\citenamefont {Dlapa},
  \citenamefont {Li},\ and\ \citenamefont {Zhang}}]{Dlapa:2021qsl}%
  \BibitemOpen
  \bibfield  {author} {\bibinfo {author} {\bibfnamefont {C.}~\bibnamefont
  {Dlapa}}, \bibinfo {author} {\bibfnamefont {X.}~\bibnamefont {Li}},\ and\
  \bibinfo {author} {\bibfnamefont {Y.}~\bibnamefont {Zhang}},\ }\href
  {https://doi.org/10.1007/JHEP07(2021)227} {\bibfield  {journal} {\bibinfo
  {journal} {JHEP}\ }\textbf {\bibinfo {volume} {07}},\ \bibinfo {pages}
  {227}},\ \Eprint {https://arxiv.org/abs/2103.04638} {arXiv:2103.04638
  [hep-th]} \BibitemShut {NoStop}%
\bibitem [{\citenamefont {He}\ \emph {et~al.}(2022)\citenamefont {He},
  \citenamefont {Li}, \citenamefont {Ma}, \citenamefont {Wu}, \citenamefont
  {Yang},\ and\ \citenamefont {Zhang}}]{He:2022ctv}%
  \BibitemOpen
  \bibfield  {author} {\bibinfo {author} {\bibfnamefont {S.}~\bibnamefont
  {He}}, \bibinfo {author} {\bibfnamefont {Z.}~\bibnamefont {Li}}, \bibinfo
  {author} {\bibfnamefont {R.}~\bibnamefont {Ma}}, \bibinfo {author}
  {\bibfnamefont {Z.}~\bibnamefont {Wu}}, \bibinfo {author} {\bibfnamefont
  {Q.}~\bibnamefont {Yang}},\ and\ \bibinfo {author} {\bibfnamefont
  {Y.}~\bibnamefont {Zhang}},\ }\href {https://doi.org/10.1007/JHEP10(2022)165}
  {\bibfield  {journal} {\bibinfo  {journal} {JHEP}\ }\textbf {\bibinfo
  {volume} {10}},\ \bibinfo {pages} {165}},\ \Eprint
  {https://arxiv.org/abs/2206.04609} {arXiv:2206.04609 [hep-th]} \BibitemShut
  {NoStop}%
\bibitem [{\citenamefont {Weinzierl}(2022)}]{Weinzierl:2022eaz}%
  \BibitemOpen
  \bibfield  {author} {\bibinfo {author} {\bibfnamefont {S.}~\bibnamefont
  {Weinzierl}},\ }\href {https://doi.org/10.1007/978-3-030-99558-4} {\emph
  {\bibinfo {title} {{Feynman Integrals}}}}\ (\bibinfo {year} {2022})\ \Eprint
  {https://arxiv.org/abs/2201.03593} {arXiv:2201.03593 [hep-th]} \BibitemShut
  {NoStop}%
\bibitem [{\citenamefont {Chicherin}\ \emph {et~al.}(2019)\citenamefont
  {Chicherin}, \citenamefont {Gehrmann}, \citenamefont {Henn}, \citenamefont
  {Wasser}, \citenamefont {Zhang},\ and\ \citenamefont
  {Zoia}}]{Chicherin:2018old}%
  \BibitemOpen
  \bibfield  {author} {\bibinfo {author} {\bibfnamefont {D.}~\bibnamefont
  {Chicherin}}, \bibinfo {author} {\bibfnamefont {T.}~\bibnamefont {Gehrmann}},
  \bibinfo {author} {\bibfnamefont {J.~M.}\ \bibnamefont {Henn}}, \bibinfo
  {author} {\bibfnamefont {P.}~\bibnamefont {Wasser}}, \bibinfo {author}
  {\bibfnamefont {Y.}~\bibnamefont {Zhang}},\ and\ \bibinfo {author}
  {\bibfnamefont {S.}~\bibnamefont {Zoia}},\ }\href
  {https://doi.org/10.1103/PhysRevLett.123.041603} {\bibfield  {journal}
  {\bibinfo  {journal} {Phys. Rev. Lett.}\ }\textbf {\bibinfo {volume} {123}},\
  \bibinfo {pages} {041603} (\bibinfo {year} {2019})},\ \Eprint
  {https://arxiv.org/abs/1812.11160} {arXiv:1812.11160 [hep-ph]} \BibitemShut
  {NoStop}%
\bibitem [{\citenamefont {Henn}\ \emph {et~al.}(2020)\citenamefont {Henn},
  \citenamefont {Mistlberger}, \citenamefont {Smirnov},\ and\ \citenamefont
  {Wasser}}]{Henn:2020lye}%
  \BibitemOpen
  \bibfield  {author} {\bibinfo {author} {\bibfnamefont {J.}~\bibnamefont
  {Henn}}, \bibinfo {author} {\bibfnamefont {B.}~\bibnamefont {Mistlberger}},
  \bibinfo {author} {\bibfnamefont {V.~A.}\ \bibnamefont {Smirnov}},\ and\
  \bibinfo {author} {\bibfnamefont {P.}~\bibnamefont {Wasser}},\ }\href
  {https://doi.org/10.1007/JHEP04(2020)167} {\bibfield  {journal} {\bibinfo
  {journal} {JHEP}\ }\textbf {\bibinfo {volume} {04}},\ \bibinfo {pages}
  {167}},\ \Eprint {https://arxiv.org/abs/2002.09492} {arXiv:2002.09492
  [hep-ph]} \BibitemShut {NoStop}%
\bibitem [{\citenamefont {Magnus}(1954)}]{magnus1954exponential}%
  \BibitemOpen
  \bibfield  {author} {\bibinfo {author} {\bibfnamefont {W.}~\bibnamefont
  {Magnus}},\ }\href {https://doi.org/10.1002/cpa.3160070404} {\bibfield
  {journal} {\bibinfo  {journal} {Commun. Pure Appl. Math.}\ }\textbf {\bibinfo
  {volume} {7}},\ \bibinfo {pages} {649} (\bibinfo {year} {1954})}\BibitemShut
  {NoStop}%
\bibitem [{\citenamefont {Blanes}\ \emph {et~al.}(2009)\citenamefont {Blanes},
  \citenamefont {Casas}, \citenamefont {Oteo},\ and\ \citenamefont
  {Ros}}]{blanes2009magnus}%
  \BibitemOpen
  \bibfield  {author} {\bibinfo {author} {\bibfnamefont {S.}~\bibnamefont
  {Blanes}}, \bibinfo {author} {\bibfnamefont {F.}~\bibnamefont {Casas}},
  \bibinfo {author} {\bibfnamefont {J.~{\'A}.}\ \bibnamefont {Oteo}},\ and\
  \bibinfo {author} {\bibfnamefont {J.}~\bibnamefont {Ros}},\ }\href
  {https://doi.org/https://doi.org/10.1016/j.physrep.2008.11.001} {\bibfield
  {journal} {\bibinfo  {journal} {Phys. Rept.}\ }\textbf {\bibinfo {volume}
  {470}},\ \bibinfo {pages} {151} (\bibinfo {year} {2009})}\BibitemShut
  {NoStop}%
\bibitem [{\citenamefont {Argeri}\ \emph {et~al.}(2014)\citenamefont {Argeri},
  \citenamefont {Di~Vita}, \citenamefont {Mastrolia}, \citenamefont
  {Mirabella}, \citenamefont {Schlenk}, \citenamefont {Schubert},\ and\
  \citenamefont {Tancredi}}]{Argeri:2014qva}%
  \BibitemOpen
  \bibfield  {author} {\bibinfo {author} {\bibfnamefont {M.}~\bibnamefont
  {Argeri}}, \bibinfo {author} {\bibfnamefont {S.}~\bibnamefont {Di~Vita}},
  \bibinfo {author} {\bibfnamefont {P.}~\bibnamefont {Mastrolia}}, \bibinfo
  {author} {\bibfnamefont {E.}~\bibnamefont {Mirabella}}, \bibinfo {author}
  {\bibfnamefont {J.}~\bibnamefont {Schlenk}}, \bibinfo {author} {\bibfnamefont
  {U.}~\bibnamefont {Schubert}},\ and\ \bibinfo {author} {\bibfnamefont
  {L.}~\bibnamefont {Tancredi}},\ }\href
  {https://doi.org/10.1007/JHEP03(2014)082} {\bibfield  {journal} {\bibinfo
  {journal} {JHEP}\ }\textbf {\bibinfo {volume} {03}},\ \bibinfo {pages}
  {082}},\ \Eprint {https://arxiv.org/abs/1401.2979} {arXiv:1401.2979 [hep-ph]}
  \BibitemShut {NoStop}%
\bibitem [{\citenamefont {Chen}(1977)}]{chen1977iterated}%
  \BibitemOpen
  \bibfield  {author} {\bibinfo {author} {\bibfnamefont {K.-T.}\ \bibnamefont
  {Chen}},\ }\href
  {https://doi.org/https://doi.org/10.1090/S0002-9904-1977-14320-6} {\bibfield
  {journal} {\bibinfo  {journal} {Bull. Amer. Math. Soc.}\ }\textbf {\bibinfo
  {volume} {83}},\ \bibinfo {pages} {831} (\bibinfo {year} {1977})}\BibitemShut
  {NoStop}%
\bibitem [{\citenamefont {Goncharov}(1998)}]{Goncharov:1998kja}%
  \BibitemOpen
  \bibfield  {author} {\bibinfo {author} {\bibfnamefont {A.~B.}\ \bibnamefont
  {Goncharov}},\ }\href {https://doi.org/10.4310/MRL.1998.v5.n4.a7} {\bibfield
  {journal} {\bibinfo  {journal} {Math. Res. Lett.}\ }\textbf {\bibinfo
  {volume} {5}},\ \bibinfo {pages} {497} (\bibinfo {year} {1998})},\ \Eprint
  {https://arxiv.org/abs/1105.2076} {arXiv:1105.2076 [math.AG]} \BibitemShut
  {NoStop}%
\bibitem [{\citenamefont {Chang}\ \emph {et~al.}(1982)\citenamefont {Chang},
  \citenamefont {Gaemers},\ and\ \citenamefont {van Neerven}}]{Chang:1981qq}%
  \BibitemOpen
  \bibfield  {author} {\bibinfo {author} {\bibfnamefont {T.~H.}\ \bibnamefont
  {Chang}}, \bibinfo {author} {\bibfnamefont {K.~J.~F.}\ \bibnamefont
  {Gaemers}},\ and\ \bibinfo {author} {\bibfnamefont {W.~L.}\ \bibnamefont {van
  Neerven}},\ }\href {https://doi.org/10.1016/0550-3213(82)90407-2} {\bibfield
  {journal} {\bibinfo  {journal} {Nucl. Phys. B}\ }\textbf {\bibinfo {volume}
  {202}},\ \bibinfo {pages} {407} (\bibinfo {year} {1982})}\BibitemShut
  {NoStop}%
\bibitem [{\citenamefont {Djouadi}\ and\ \citenamefont
  {Verzegnassi}(1987)}]{Djouadi:1987gn}%
  \BibitemOpen
  \bibfield  {author} {\bibinfo {author} {\bibfnamefont {A.}~\bibnamefont
  {Djouadi}}\ and\ \bibinfo {author} {\bibfnamefont {C.}~\bibnamefont
  {Verzegnassi}},\ }\href {https://doi.org/10.1016/0370-2693(87)91206-8}
  {\bibfield  {journal} {\bibinfo  {journal} {Phys. Lett. B}\ }\textbf
  {\bibinfo {volume} {195}},\ \bibinfo {pages} {265} (\bibinfo {year}
  {1987})}\BibitemShut {NoStop}%
\bibitem [{\citenamefont {Djouadi}(1988)}]{Djouadi:1987di}%
  \BibitemOpen
  \bibfield  {author} {\bibinfo {author} {\bibfnamefont {A.}~\bibnamefont
  {Djouadi}},\ }\href {https://doi.org/10.1007/BF02812964} {\bibfield
  {journal} {\bibinfo  {journal} {Nuovo Cim. A}\ }\textbf {\bibinfo {volume}
  {100}},\ \bibinfo {pages} {357} (\bibinfo {year} {1988})}\BibitemShut
  {NoStop}%
\bibitem [{\citenamefont {Kniehl}\ \emph {et~al.}(1988)\citenamefont {Kniehl},
  \citenamefont {K{\"u}hn},\ and\ \citenamefont {Stuart}}]{Kniehl:1988ie}%
  \BibitemOpen
  \bibfield  {author} {\bibinfo {author} {\bibfnamefont {B.~A.}\ \bibnamefont
  {Kniehl}}, \bibinfo {author} {\bibfnamefont {J.~H.}\ \bibnamefont
  {K{\"u}hn}},\ and\ \bibinfo {author} {\bibfnamefont {R.~G.}\ \bibnamefont
  {Stuart}},\ }\href {https://doi.org/10.1016/0370-2693(88)90132-3} {\bibfield
  {journal} {\bibinfo  {journal} {Phys. Lett. B}\ }\textbf {\bibinfo {volume}
  {214}},\ \bibinfo {pages} {621} (\bibinfo {year} {1988})}\BibitemShut
  {NoStop}%
\bibitem [{\citenamefont {Kniehl}(1990)}]{Kniehl:1989yc}%
  \BibitemOpen
  \bibfield  {author} {\bibinfo {author} {\bibfnamefont {B.~A.}\ \bibnamefont
  {Kniehl}},\ }\href {https://doi.org/10.1016/0550-3213(90)90552-O} {\bibfield
  {journal} {\bibinfo  {journal} {Nucl. Phys. B}\ }\textbf {\bibinfo {volume}
  {347}},\ \bibinfo {pages} {86} (\bibinfo {year} {1990})}\BibitemShut
  {NoStop}%
\bibitem [{\citenamefont {Djouadi}\ and\ \citenamefont
  {Gambino}(1994)}]{Djouadi:1993ss}%
  \BibitemOpen
  \bibfield  {author} {\bibinfo {author} {\bibfnamefont {A.}~\bibnamefont
  {Djouadi}}\ and\ \bibinfo {author} {\bibfnamefont {P.}~\bibnamefont
  {Gambino}},\ }\href {https://doi.org/10.1103/PhysRevD.49.3499} {\bibfield
  {journal} {\bibinfo  {journal} {Phys. Rev. D}\ }\textbf {\bibinfo {volume}
  {49}},\ \bibinfo {pages} {3499} (\bibinfo {year} {1994})},\ \bibinfo {note}
  {[Erratum: Phys.Rev.D 53, 4111 (1996)]},\ \Eprint
  {https://arxiv.org/abs/hep-ph/9309298} {arXiv:hep-ph/9309298} \BibitemShut
  {NoStop}%
\bibitem [{\citenamefont {Usyukina}\ and\ \citenamefont
  {Davydychev}(1994)}]{Usyukina:1994iw}%
  \BibitemOpen
  \bibfield  {author} {\bibinfo {author} {\bibfnamefont {N.~I.}\ \bibnamefont
  {Usyukina}}\ and\ \bibinfo {author} {\bibfnamefont {A.~I.}\ \bibnamefont
  {Davydychev}},\ }\href {https://doi.org/10.1016/0370-2693(94)90874-5}
  {\bibfield  {journal} {\bibinfo  {journal} {Phys. Lett. B}\ }\textbf
  {\bibinfo {volume} {332}},\ \bibinfo {pages} {159} (\bibinfo {year}
  {1994})},\ \Eprint {https://arxiv.org/abs/hep-ph/9402223}
  {arXiv:hep-ph/9402223} \BibitemShut {NoStop}%
\bibitem [{\citenamefont {Birthwright}\ \emph {et~al.}(2004)\citenamefont
  {Birthwright}, \citenamefont {Glover},\ and\ \citenamefont
  {Marquard}}]{Birthwright:2004kk}%
  \BibitemOpen
  \bibfield  {author} {\bibinfo {author} {\bibfnamefont {T.~G.}\ \bibnamefont
  {Birthwright}}, \bibinfo {author} {\bibfnamefont {E.~W.~N.}\ \bibnamefont
  {Glover}},\ and\ \bibinfo {author} {\bibfnamefont {P.}~\bibnamefont
  {Marquard}},\ }\href {https://doi.org/10.1088/1126-6708/2004/09/042}
  {\bibfield  {journal} {\bibinfo  {journal} {JHEP}\ }\textbf {\bibinfo
  {volume} {09}},\ \bibinfo {pages} {042}},\ \Eprint
  {https://arxiv.org/abs/hep-ph/0407343} {arXiv:hep-ph/0407343} \BibitemShut
  {NoStop}%
\bibitem [{\citenamefont {Chavez}\ and\ \citenamefont
  {Duhr}(2012)}]{Chavez:2012kn}%
  \BibitemOpen
  \bibfield  {author} {\bibinfo {author} {\bibfnamefont {F.}~\bibnamefont
  {Chavez}}\ and\ \bibinfo {author} {\bibfnamefont {C.}~\bibnamefont {Duhr}},\
  }\href {https://doi.org/10.1007/JHEP11(2012)114} {\bibfield  {journal}
  {\bibinfo  {journal} {JHEP}\ }\textbf {\bibinfo {volume} {11}},\ \bibinfo
  {pages} {114}},\ \Eprint {https://arxiv.org/abs/1209.2722} {arXiv:1209.2722
  [hep-ph]} \BibitemShut {NoStop}%
\bibitem [{\citenamefont {Fleischer}\ \emph {et~al.}(2004)\citenamefont
  {Fleischer}, \citenamefont {Tarasov},\ and\ \citenamefont
  {Tarasov}}]{Fleischer:2004vb}%
  \BibitemOpen
  \bibfield  {author} {\bibinfo {author} {\bibfnamefont {J.}~\bibnamefont
  {Fleischer}}, \bibinfo {author} {\bibfnamefont {O.~V.}\ \bibnamefont
  {Tarasov}},\ and\ \bibinfo {author} {\bibfnamefont {V.~O.}\ \bibnamefont
  {Tarasov}},\ }\href {https://doi.org/10.1016/j.physletb.2004.01.063}
  {\bibfield  {journal} {\bibinfo  {journal} {Phys. Lett. B}\ }\textbf
  {\bibinfo {volume} {584}},\ \bibinfo {pages} {294} (\bibinfo {year}
  {2004})},\ \Eprint {https://arxiv.org/abs/hep-ph/0401090}
  {arXiv:hep-ph/0401090} \BibitemShut {NoStop}%
\bibitem [{\citenamefont {Harlander}\ and\ \citenamefont
  {Kant}(2005)}]{Harlander:2005rq}%
  \BibitemOpen
  \bibfield  {author} {\bibinfo {author} {\bibfnamefont {R.}~\bibnamefont
  {Harlander}}\ and\ \bibinfo {author} {\bibfnamefont {P.}~\bibnamefont
  {Kant}},\ }\href {https://doi.org/10.1088/1126-6708/2005/12/015} {\bibfield
  {journal} {\bibinfo  {journal} {JHEP}\ }\textbf {\bibinfo {volume} {12}},\
  \bibinfo {pages} {015}},\ \Eprint {https://arxiv.org/abs/hep-ph/0509189}
  {arXiv:hep-ph/0509189} \BibitemShut {NoStop}%
\bibitem [{\citenamefont {Aglietti}\ \emph {et~al.}(2007)\citenamefont
  {Aglietti}, \citenamefont {Bonciani}, \citenamefont {Degrassi},\ and\
  \citenamefont {Vicini}}]{Aglietti:2006tp}%
  \BibitemOpen
  \bibfield  {author} {\bibinfo {author} {\bibfnamefont {U.}~\bibnamefont
  {Aglietti}}, \bibinfo {author} {\bibfnamefont {R.}~\bibnamefont {Bonciani}},
  \bibinfo {author} {\bibfnamefont {G.}~\bibnamefont {Degrassi}},\ and\
  \bibinfo {author} {\bibfnamefont {A.}~\bibnamefont {Vicini}},\ }\href
  {https://doi.org/10.1088/1126-6708/2007/01/021} {\bibfield  {journal}
  {\bibinfo  {journal} {JHEP}\ }\textbf {\bibinfo {volume} {01}},\ \bibinfo
  {pages} {021}},\ \Eprint {https://arxiv.org/abs/hep-ph/0611266}
  {arXiv:hep-ph/0611266} \BibitemShut {NoStop}%
\bibitem [{\citenamefont {Bonciani}\ \emph {et~al.}(2015)\citenamefont
  {Bonciani}, \citenamefont {Del~Duca}, \citenamefont {Frellesvig},
  \citenamefont {Henn}, \citenamefont {Moriello},\ and\ \citenamefont
  {Smirnov}}]{Bonciani:2015eua}%
  \BibitemOpen
  \bibfield  {author} {\bibinfo {author} {\bibfnamefont {R.}~\bibnamefont
  {Bonciani}}, \bibinfo {author} {\bibfnamefont {V.}~\bibnamefont {Del~Duca}},
  \bibinfo {author} {\bibfnamefont {H.}~\bibnamefont {Frellesvig}}, \bibinfo
  {author} {\bibfnamefont {J.~M.}\ \bibnamefont {Henn}}, \bibinfo {author}
  {\bibfnamefont {F.}~\bibnamefont {Moriello}},\ and\ \bibinfo {author}
  {\bibfnamefont {V.~A.}\ \bibnamefont {Smirnov}},\ }\href
  {https://doi.org/10.1007/JHEP08(2015)108} {\bibfield  {journal} {\bibinfo
  {journal} {JHEP}\ }\textbf {\bibinfo {volume} {08}},\ \bibinfo {pages}
  {108}},\ \Eprint {https://arxiv.org/abs/1505.00567} {arXiv:1505.00567
  [hep-ph]} \BibitemShut {NoStop}%
\bibitem [{\citenamefont {Gehrmann}\ \emph {et~al.}(2015)\citenamefont
  {Gehrmann}, \citenamefont {Guns},\ and\ \citenamefont
  {Kara}}]{Gehrmann:2015dua}%
  \BibitemOpen
  \bibfield  {author} {\bibinfo {author} {\bibfnamefont {T.}~\bibnamefont
  {Gehrmann}}, \bibinfo {author} {\bibfnamefont {S.}~\bibnamefont {Guns}},\
  and\ \bibinfo {author} {\bibfnamefont {D.}~\bibnamefont {Kara}},\ }\href
  {https://doi.org/10.1007/JHEP09(2015)038} {\bibfield  {journal} {\bibinfo
  {journal} {JHEP}\ }\textbf {\bibinfo {volume} {09}},\ \bibinfo {pages}
  {038}},\ \Eprint {https://arxiv.org/abs/1505.00561} {arXiv:1505.00561
  [hep-ph]} \BibitemShut {NoStop}%
\bibitem [{\citenamefont {Di~Vita}\ \emph {et~al.}(2017)\citenamefont
  {Di~Vita}, \citenamefont {Mastrolia}, \citenamefont {Primo},\ and\
  \citenamefont {Schubert}}]{DiVita:2017xlr}%
  \BibitemOpen
  \bibfield  {author} {\bibinfo {author} {\bibfnamefont {S.}~\bibnamefont
  {Di~Vita}}, \bibinfo {author} {\bibfnamefont {P.}~\bibnamefont {Mastrolia}},
  \bibinfo {author} {\bibfnamefont {A.}~\bibnamefont {Primo}},\ and\ \bibinfo
  {author} {\bibfnamefont {U.}~\bibnamefont {Schubert}},\ }\href
  {https://doi.org/10.1007/JHEP04(2017)008} {\bibfield  {journal} {\bibinfo
  {journal} {JHEP}\ }\textbf {\bibinfo {volume} {04}},\ \bibinfo {pages}
  {008}},\ \Eprint {https://arxiv.org/abs/1702.07331} {arXiv:1702.07331
  [hep-ph]} \BibitemShut {NoStop}%
\bibitem [{\citenamefont {Ma}\ \emph {et~al.}(2021)\citenamefont {Ma},
  \citenamefont {Wang}, \citenamefont {Xu}, \citenamefont {Yang},\ and\
  \citenamefont {Zhou}}]{Ma:2021cxg}%
  \BibitemOpen
  \bibfield  {author} {\bibinfo {author} {\bibfnamefont {C.}~\bibnamefont
  {Ma}}, \bibinfo {author} {\bibfnamefont {Y.}~\bibnamefont {Wang}}, \bibinfo
  {author} {\bibfnamefont {X.}~\bibnamefont {Xu}}, \bibinfo {author}
  {\bibfnamefont {L.~L.}\ \bibnamefont {Yang}},\ and\ \bibinfo {author}
  {\bibfnamefont {B.}~\bibnamefont {Zhou}},\ }\href
  {https://doi.org/10.1007/JHEP09(2021)114} {\bibfield  {journal} {\bibinfo
  {journal} {JHEP}\ }\textbf {\bibinfo {volume} {09}},\ \bibinfo {pages}
  {114}},\ \Eprint {https://arxiv.org/abs/2105.06316} {arXiv:2105.06316
  [hep-ph]} \BibitemShut {NoStop}%
\bibitem [{\citenamefont {Li}\ \emph {et~al.}(2024)\citenamefont {Li},
  \citenamefont {Zhang}, \citenamefont {Li}, \citenamefont {Wang},
  \citenamefont {He}, \citenamefont {Han}, \citenamefont {Jiang},\ and\
  \citenamefont {Wang}}]{Li:2024dlh}%
  \BibitemOpen
  \bibfield  {author} {\bibinfo {author} {\bibfnamefont {Z.}~\bibnamefont
  {Li}}, \bibinfo {author} {\bibfnamefont {R.-Y.}\ \bibnamefont {Zhang}},
  \bibinfo {author} {\bibfnamefont {S.-X.}\ \bibnamefont {Li}}, \bibinfo
  {author} {\bibfnamefont {X.-F.}\ \bibnamefont {Wang}}, \bibinfo {author}
  {\bibfnamefont {W.-J.}\ \bibnamefont {He}}, \bibinfo {author} {\bibfnamefont
  {L.}~\bibnamefont {Han}}, \bibinfo {author} {\bibfnamefont {Y.}~\bibnamefont
  {Jiang}},\ and\ \bibinfo {author} {\bibfnamefont {Q.-h.}\ \bibnamefont
  {Wang}},\ }\href@noop {} {\  (\bibinfo {year} {2024})},\ \Eprint
  {https://arxiv.org/abs/2401.01323} {arXiv:2401.01323 [hep-ph]} \BibitemShut
  {NoStop}%
\bibitem [{\citenamefont {Wang}\ \emph {et~al.}(2019)\citenamefont {Wang},
  \citenamefont {Xu},\ and\ \citenamefont {Yang}}]{Wang:2019fxh}%
  \BibitemOpen
  \bibfield  {author} {\bibinfo {author} {\bibfnamefont {Y.}~\bibnamefont
  {Wang}}, \bibinfo {author} {\bibfnamefont {X.}~\bibnamefont {Xu}},\ and\
  \bibinfo {author} {\bibfnamefont {L.~L.}\ \bibnamefont {Yang}},\ }\href
  {https://doi.org/10.1103/PhysRevD.100.071502} {\bibfield  {journal} {\bibinfo
   {journal} {Phys. Rev. D}\ }\textbf {\bibinfo {volume} {100}},\ \bibinfo
  {pages} {071502} (\bibinfo {year} {2019})},\ \Eprint
  {https://arxiv.org/abs/1905.11463} {arXiv:1905.11463 [hep-ph]} \BibitemShut
  {NoStop}%
\bibitem [{\citenamefont {Chaubey}\ \emph {et~al.}(2022)\citenamefont
  {Chaubey}, \citenamefont {Kaur},\ and\ \citenamefont
  {Shivaji}}]{Chaubey:2022hlr}%
  \BibitemOpen
  \bibfield  {author} {\bibinfo {author} {\bibfnamefont {E.}~\bibnamefont
  {Chaubey}}, \bibinfo {author} {\bibfnamefont {M.}~\bibnamefont {Kaur}},\ and\
  \bibinfo {author} {\bibfnamefont {A.}~\bibnamefont {Shivaji}},\ }\href
  {https://doi.org/10.1007/JHEP10(2022)056} {\bibfield  {journal} {\bibinfo
  {journal} {JHEP}\ }\textbf {\bibinfo {volume} {10}},\ \bibinfo {pages}
  {056}},\ \Eprint {https://arxiv.org/abs/2205.06339} {arXiv:2205.06339
  [hep-ph]} \BibitemShut {NoStop}%
\bibitem [{\citenamefont {Denner}(1993)}]{Denner:1991kt}%
  \BibitemOpen
  \bibfield  {author} {\bibinfo {author} {\bibfnamefont {A.}~\bibnamefont
  {Denner}},\ }\href {https://doi.org/10.1002/prop.2190410402} {\bibfield
  {journal} {\bibinfo  {journal} {Fortsch. Phys.}\ }\textbf {\bibinfo {volume}
  {41}},\ \bibinfo {pages} {307} (\bibinfo {year} {1993})},\ \Eprint
  {https://arxiv.org/abs/0709.1075} {arXiv:0709.1075 [hep-ph]} \BibitemShut
  {NoStop}%
\bibitem [{\citenamefont {Denner}\ and\ \citenamefont
  {Dittmaier}(2020)}]{Denner:2019vbn}%
  \BibitemOpen
  \bibfield  {author} {\bibinfo {author} {\bibfnamefont {A.}~\bibnamefont
  {Denner}}\ and\ \bibinfo {author} {\bibfnamefont {S.}~\bibnamefont
  {Dittmaier}},\ }\href {https://doi.org/10.1016/j.physrep.2020.04.001}
  {\bibfield  {journal} {\bibinfo  {journal} {Phys. Rept.}\ }\textbf {\bibinfo
  {volume} {864}},\ \bibinfo {pages} {1} (\bibinfo {year} {2020})},\ \Eprint
  {https://arxiv.org/abs/1912.06823} {arXiv:1912.06823 [hep-ph]} \BibitemShut
  {NoStop}%
\bibitem [{\citenamefont {'t~Hooft}\ and\ \citenamefont
  {Veltman}(1972)}]{tHooft:1972tcz}%
  \BibitemOpen
  \bibfield  {author} {\bibinfo {author} {\bibfnamefont {G.}~\bibnamefont
  {'t~Hooft}}\ and\ \bibinfo {author} {\bibfnamefont {M.~J.~G.}\ \bibnamefont
  {Veltman}},\ }\href {https://doi.org/10.1016/0550-3213(72)90279-9} {\bibfield
   {journal} {\bibinfo  {journal} {Nucl. Phys. B}\ }\textbf {\bibinfo {volume}
  {44}},\ \bibinfo {pages} {189} (\bibinfo {year} {1972})}\BibitemShut
  {NoStop}%
\bibitem [{\citenamefont {Bollini}\ and\ \citenamefont
  {Giambiagi}(1972)}]{Bollini:1972ui}%
  \BibitemOpen
  \bibfield  {author} {\bibinfo {author} {\bibfnamefont {C.~G.}\ \bibnamefont
  {Bollini}}\ and\ \bibinfo {author} {\bibfnamefont {J.~J.}\ \bibnamefont
  {Giambiagi}},\ }\href {https://doi.org/10.1007/BF02895558} {\bibfield
  {journal} {\bibinfo  {journal} {Nuovo Cim. B}\ }\textbf {\bibinfo {volume}
  {12}},\ \bibinfo {pages} {20} (\bibinfo {year} {1972})}\BibitemShut {NoStop}%
\bibitem [{\citenamefont {Dittmaier}(2000)}]{Dittmaier:1999mb}%
  \BibitemOpen
  \bibfield  {author} {\bibinfo {author} {\bibfnamefont {S.}~\bibnamefont
  {Dittmaier}},\ }\href {https://doi.org/10.1016/S0550-3213(99)00563-5}
  {\bibfield  {journal} {\bibinfo  {journal} {Nucl. Phys. B}\ }\textbf
  {\bibinfo {volume} {565}},\ \bibinfo {pages} {69} (\bibinfo {year} {2000})},\
  \Eprint {https://arxiv.org/abs/hep-ph/9904440} {arXiv:hep-ph/9904440}
  \BibitemShut {NoStop}%
\bibitem [{\citenamefont {Denner}\ \emph {et~al.}(2000)\citenamefont {Denner},
  \citenamefont {Dittmaier}, \citenamefont {Roth},\ and\ \citenamefont
  {Wackeroth}}]{Denner:2000bj}%
  \BibitemOpen
  \bibfield  {author} {\bibinfo {author} {\bibfnamefont {A.}~\bibnamefont
  {Denner}}, \bibinfo {author} {\bibfnamefont {S.}~\bibnamefont {Dittmaier}},
  \bibinfo {author} {\bibfnamefont {M.}~\bibnamefont {Roth}},\ and\ \bibinfo
  {author} {\bibfnamefont {D.}~\bibnamefont {Wackeroth}},\ }\href
  {https://doi.org/10.1016/S0550-3213(00)00511-3} {\bibfield  {journal}
  {\bibinfo  {journal} {Nucl. Phys. B}\ }\textbf {\bibinfo {volume} {587}},\
  \bibinfo {pages} {67} (\bibinfo {year} {2000})},\ \Eprint
  {https://arxiv.org/abs/hep-ph/0006307} {arXiv:hep-ph/0006307} \BibitemShut
  {NoStop}%
\bibitem [{\citenamefont {Harris}\ and\ \citenamefont
  {Owens}(2002)}]{Harris:2001sx}%
  \BibitemOpen
  \bibfield  {author} {\bibinfo {author} {\bibfnamefont {B.~W.}\ \bibnamefont
  {Harris}}\ and\ \bibinfo {author} {\bibfnamefont {J.~F.}\ \bibnamefont
  {Owens}},\ }\href {https://doi.org/10.1103/PhysRevD.65.094032} {\bibfield
  {journal} {\bibinfo  {journal} {Phys. Rev. D}\ }\textbf {\bibinfo {volume}
  {65}},\ \bibinfo {pages} {094032} (\bibinfo {year} {2002})},\ \Eprint
  {https://arxiv.org/abs/hep-ph/0102128} {arXiv:hep-ph/0102128} \BibitemShut
  {NoStop}%
\bibitem [{\citenamefont {Beenakker}\ \emph {et~al.}(1996)\citenamefont
  {Beenakker}, \citenamefont {Berends}, \citenamefont {Argyres}, \citenamefont
  {Bardin}, \citenamefont {Denner}, \citenamefont {Dittmaier} \emph
  {et~al.}}]{Beenakker:1996kt}%
  \BibitemOpen
  \bibfield  {author} {\bibinfo {author} {\bibfnamefont {W.}~\bibnamefont
  {Beenakker}}, \bibinfo {author} {\bibfnamefont {F.~A.}\ \bibnamefont
  {Berends}}, \bibinfo {author} {\bibfnamefont {E.~N.}\ \bibnamefont
  {Argyres}}, \bibinfo {author} {\bibfnamefont {D.~Y.}\ \bibnamefont {Bardin}},
  \bibinfo {author} {\bibfnamefont {A.}~\bibnamefont {Denner}}, \bibinfo
  {author} {\bibfnamefont {S.}~\bibnamefont {Dittmaier}}, \emph {et~al.},\
  }\bibfield  {journal} {\bibinfo  {journal} {CERN Yellow Reports: Workshop on
  LEP2 Physics}\ }\href {https://doi.org/10.5170/CERN-1996-001-V-1.79}
  {10.5170/CERN-1996-001-V-1.79} (\bibinfo {year} {1996}),\ \Eprint
  {https://arxiv.org/abs/hep-ph/9602351} {arXiv:hep-ph/9602351} \BibitemShut
  {NoStop}%
\bibitem [{\citenamefont {Sirlin}(1980)}]{Sirlin:1980nh}%
  \BibitemOpen
  \bibfield  {author} {\bibinfo {author} {\bibfnamefont {A.}~\bibnamefont
  {Sirlin}},\ }\href {https://doi.org/10.1103/PhysRevD.22.971} {\bibfield
  {journal} {\bibinfo  {journal} {Phys. Rev. D}\ }\textbf {\bibinfo {volume}
  {22}},\ \bibinfo {pages} {971} (\bibinfo {year} {1980})}\BibitemShut
  {NoStop}%
\bibitem [{\citenamefont {Hahn}\ and\ \citenamefont
  {P{\'e}rez-Victoria}(1999)}]{Hahn:1998yk}%
  \BibitemOpen
  \bibfield  {author} {\bibinfo {author} {\bibfnamefont {T.}~\bibnamefont
  {Hahn}}\ and\ \bibinfo {author} {\bibfnamefont {M.}~\bibnamefont
  {P{\'e}rez-Victoria}},\ }\href
  {https://doi.org/10.1016/S0010-4655(98)00173-8} {\bibfield  {journal}
  {\bibinfo  {journal} {Comput. Phys. Commun.}\ }\textbf {\bibinfo {volume}
  {118}},\ \bibinfo {pages} {153} (\bibinfo {year} {1999})},\ \Eprint
  {https://arxiv.org/abs/hep-ph/9807565} {arXiv:hep-ph/9807565} \BibitemShut
  {NoStop}%
\bibitem [{\citenamefont {van Oldenborgh}(1991)}]{vanOldenborgh:1990yc}%
  \BibitemOpen
  \bibfield  {author} {\bibinfo {author} {\bibfnamefont {G.~J.}\ \bibnamefont
  {van Oldenborgh}},\ }\href {https://doi.org/10.1016/0010-4655(91)90002-3}
  {\bibfield  {journal} {\bibinfo  {journal} {Comput. Phys. Commun.}\ }\textbf
  {\bibinfo {volume} {66}},\ \bibinfo {pages} {1} (\bibinfo {year}
  {1991})}\BibitemShut {NoStop}%
\bibitem [{\citenamefont {Breitenlohner}\ and\ \citenamefont
  {Maison}(1977)}]{Breitenlohner:1977hr}%
  \BibitemOpen
  \bibfield  {author} {\bibinfo {author} {\bibfnamefont {P.}~\bibnamefont
  {Breitenlohner}}\ and\ \bibinfo {author} {\bibfnamefont {D.}~\bibnamefont
  {Maison}},\ }\href {https://doi.org/10.1007/BF01609069} {\bibfield  {journal}
  {\bibinfo  {journal} {Commun. Math. Phys.}\ }\textbf {\bibinfo {volume}
  {52}},\ \bibinfo {pages} {11} (\bibinfo {year} {1977})}\BibitemShut {NoStop}%
\bibitem [{\citenamefont {Aoyama}\ and\ \citenamefont
  {Tonin}(1981)}]{Aoyama:1980yw}%
  \BibitemOpen
  \bibfield  {author} {\bibinfo {author} {\bibfnamefont {S.}~\bibnamefont
  {Aoyama}}\ and\ \bibinfo {author} {\bibfnamefont {M.}~\bibnamefont {Tonin}},\
  }\href {https://doi.org/10.1016/0550-3213(81)90240-6} {\bibfield  {journal}
  {\bibinfo  {journal} {Nucl. Phys. B}\ }\textbf {\bibinfo {volume} {179}},\
  \bibinfo {pages} {293} (\bibinfo {year} {1981})}\BibitemShut {NoStop}%
\bibitem [{\citenamefont {Bonneau}(1990)}]{Bonneau:1990xu}%
  \BibitemOpen
  \bibfield  {author} {\bibinfo {author} {\bibfnamefont {G.}~\bibnamefont
  {Bonneau}},\ }\href {https://doi.org/10.1142/S0217751X90001641} {\bibfield
  {journal} {\bibinfo  {journal} {Int. J. Mod. Phys. A}\ }\textbf {\bibinfo
  {volume} {5}},\ \bibinfo {pages} {3831} (\bibinfo {year} {1990})}\BibitemShut
  {NoStop}%
\bibitem [{\citenamefont {Barroso}\ \emph {et~al.}(1991)\citenamefont
  {Barroso}, \citenamefont {Doncheski}, \citenamefont {Grotch}, \citenamefont
  {K{\"o}rner},\ and\ \citenamefont {Schilcher}}]{Barroso:1990ti}%
  \BibitemOpen
  \bibfield  {author} {\bibinfo {author} {\bibfnamefont {A.}~\bibnamefont
  {Barroso}}, \bibinfo {author} {\bibfnamefont {M.~A.}\ \bibnamefont
  {Doncheski}}, \bibinfo {author} {\bibfnamefont {H.}~\bibnamefont {Grotch}},
  \bibinfo {author} {\bibfnamefont {J.~G.}\ \bibnamefont {K{\"o}rner}},\ and\
  \bibinfo {author} {\bibfnamefont {K.}~\bibnamefont {Schilcher}},\ }\href
  {https://doi.org/10.1016/0370-2693(91)91336-T} {\bibfield  {journal}
  {\bibinfo  {journal} {Phys. Lett. B}\ }\textbf {\bibinfo {volume} {261}},\
  \bibinfo {pages} {123} (\bibinfo {year} {1991})}\BibitemShut {NoStop}%
\bibitem [{\citenamefont {Larin}(1993)}]{Larin:1993tq}%
  \BibitemOpen
  \bibfield  {author} {\bibinfo {author} {\bibfnamefont {S.~A.}\ \bibnamefont
  {Larin}},\ }\href {https://doi.org/10.1016/0370-2693(93)90053-K} {\bibfield
  {journal} {\bibinfo  {journal} {Phys. Lett. B}\ }\textbf {\bibinfo {volume}
  {303}},\ \bibinfo {pages} {113} (\bibinfo {year} {1993})},\ \Eprint
  {https://arxiv.org/abs/hep-ph/9302240} {arXiv:hep-ph/9302240} \BibitemShut
  {NoStop}%
\bibitem [{\citenamefont {Kreimer}(1990)}]{Kreimer:1989ke}%
  \BibitemOpen
  \bibfield  {author} {\bibinfo {author} {\bibfnamefont {D.}~\bibnamefont
  {Kreimer}},\ }\href {https://doi.org/10.1016/0370-2693(90)90461-E} {\bibfield
   {journal} {\bibinfo  {journal} {Phys. Lett. B}\ }\textbf {\bibinfo {volume}
  {237}},\ \bibinfo {pages} {59} (\bibinfo {year} {1990})}\BibitemShut
  {NoStop}%
\bibitem [{\citenamefont {K{\"o}rner}\ \emph {et~al.}(1992)\citenamefont
  {K{\"o}rner}, \citenamefont {Kreimer},\ and\ \citenamefont
  {Schilcher}}]{Korner:1991sx}%
  \BibitemOpen
  \bibfield  {author} {\bibinfo {author} {\bibfnamefont {J.~G.}\ \bibnamefont
  {K{\"o}rner}}, \bibinfo {author} {\bibfnamefont {D.}~\bibnamefont
  {Kreimer}},\ and\ \bibinfo {author} {\bibfnamefont {K.}~\bibnamefont
  {Schilcher}},\ }\href {https://doi.org/10.1007/BF01559471} {\bibfield
  {journal} {\bibinfo  {journal} {Z. Phys. C}\ }\textbf {\bibinfo {volume}
  {54}},\ \bibinfo {pages} {503} (\bibinfo {year} {1992})}\BibitemShut
  {NoStop}%
\bibitem [{\citenamefont {Kreimer}(1993)}]{Kreimer:1993bh}%
  \BibitemOpen
  \bibfield  {author} {\bibinfo {author} {\bibfnamefont {D.}~\bibnamefont
  {Kreimer}},\ }\href@noop {} {\  (\bibinfo {year} {1993})},\ \Eprint
  {https://arxiv.org/abs/hep-ph/9401354} {arXiv:hep-ph/9401354} \BibitemShut
  {NoStop}%
\bibitem [{\citenamefont {Chen}\ \emph {et~al.}(2024)\citenamefont {Chen},
  \citenamefont {Guan}, \citenamefont {He}, \citenamefont {Liu},\ and\
  \citenamefont {Ma}}]{Chen:2022vzo}%
  \BibitemOpen
  \bibfield  {author} {\bibinfo {author} {\bibfnamefont {X.}~\bibnamefont
  {Chen}}, \bibinfo {author} {\bibfnamefont {X.}~\bibnamefont {Guan}}, \bibinfo
  {author} {\bibfnamefont {C.-Q.}\ \bibnamefont {He}}, \bibinfo {author}
  {\bibfnamefont {X.}~\bibnamefont {Liu}},\ and\ \bibinfo {author}
  {\bibfnamefont {Y.-Q.}\ \bibnamefont {Ma}},\ }\href
  {https://doi.org/10.1103/PhysRevLett.132.101901} {\bibfield  {journal}
  {\bibinfo  {journal} {Phys. Rev. Lett.}\ }\textbf {\bibinfo {volume} {132}},\
  \bibinfo {pages} {101901} (\bibinfo {year} {2024})},\ \Eprint
  {https://arxiv.org/abs/2209.14259} {arXiv:2209.14259 [hep-ph]} \BibitemShut
  {NoStop}%
\bibitem [{\citenamefont {Chen}(2023)}]{Chen:2023lus}%
  \BibitemOpen
  \bibfield  {author} {\bibinfo {author} {\bibfnamefont {L.}~\bibnamefont
  {Chen}},\ }\href {https://doi.org/10.1007/JHEP11(2023)030} {\bibfield
  {journal} {\bibinfo  {journal} {JHEP}\ }\textbf {\bibinfo {volume} {2023}},\
  \bibinfo {pages} {30}},\ \Eprint {https://arxiv.org/abs/2304.13814}
  {arXiv:2304.13814 [hep-ph]} \BibitemShut {NoStop}%
\bibitem [{\citenamefont {Jegerlehner}(2001)}]{Jegerlehner:2000dz}%
  \BibitemOpen
  \bibfield  {author} {\bibinfo {author} {\bibfnamefont {F.}~\bibnamefont
  {Jegerlehner}},\ }\href {https://doi.org/10.1007/s100520100573} {\bibfield
  {journal} {\bibinfo  {journal} {Eur. Phys. J. C}\ }\textbf {\bibinfo {volume}
  {18}},\ \bibinfo {pages} {673} (\bibinfo {year} {2001})},\ \Eprint
  {https://arxiv.org/abs/hep-th/0005255} {arXiv:hep-th/0005255} \BibitemShut
  {NoStop}%
\bibitem [{\citenamefont {Bernreuther}\ \emph {et~al.}(2005)\citenamefont
  {Bernreuther}, \citenamefont {Bonciani}, \citenamefont {Gehrmann},
  \citenamefont {Heinesch}, \citenamefont {Leineweber}, \citenamefont
  {Mastrolia},\ and\ \citenamefont {Remiddi}}]{Bernreuther:2004ih}%
  \BibitemOpen
  \bibfield  {author} {\bibinfo {author} {\bibfnamefont {W.}~\bibnamefont
  {Bernreuther}}, \bibinfo {author} {\bibfnamefont {R.}~\bibnamefont
  {Bonciani}}, \bibinfo {author} {\bibfnamefont {T.}~\bibnamefont {Gehrmann}},
  \bibinfo {author} {\bibfnamefont {R.}~\bibnamefont {Heinesch}}, \bibinfo
  {author} {\bibfnamefont {T.}~\bibnamefont {Leineweber}}, \bibinfo {author}
  {\bibfnamefont {P.}~\bibnamefont {Mastrolia}},\ and\ \bibinfo {author}
  {\bibfnamefont {E.}~\bibnamefont {Remiddi}},\ }\href
  {https://doi.org/10.1016/j.nuclphysb.2004.10.059} {\bibfield  {journal}
  {\bibinfo  {journal} {Nucl. Phys. B}\ }\textbf {\bibinfo {volume} {706}},\
  \bibinfo {pages} {245} (\bibinfo {year} {2005})},\ \Eprint
  {https://arxiv.org/abs/hep-ph/0406046} {arXiv:hep-ph/0406046} \BibitemShut
  {NoStop}%
\bibitem [{\citenamefont {Dittmaier}\ \emph {et~al.}(2016)\citenamefont
  {Dittmaier}, \citenamefont {Huss},\ and\ \citenamefont
  {Schwinn}}]{Dittmaier:2015rxo}%
  \BibitemOpen
  \bibfield  {author} {\bibinfo {author} {\bibfnamefont {S.}~\bibnamefont
  {Dittmaier}}, \bibinfo {author} {\bibfnamefont {A.}~\bibnamefont {Huss}},\
  and\ \bibinfo {author} {\bibfnamefont {C.}~\bibnamefont {Schwinn}},\ }\href
  {https://doi.org/10.1016/j.nuclphysb.2016.01.006} {\bibfield  {journal}
  {\bibinfo  {journal} {Nucl. Phys. B}\ }\textbf {\bibinfo {volume} {904}},\
  \bibinfo {pages} {216} (\bibinfo {year} {2016})},\ \Eprint
  {https://arxiv.org/abs/1511.08016} {arXiv:1511.08016 [hep-ph]} \BibitemShut
  {NoStop}%
\bibitem [{\citenamefont {Hahn}(2001)}]{Hahn:2000kx}%
  \BibitemOpen
  \bibfield  {author} {\bibinfo {author} {\bibfnamefont {T.}~\bibnamefont
  {Hahn}},\ }\href {https://doi.org/10.1016/S0010-4655(01)00290-9} {\bibfield
  {journal} {\bibinfo  {journal} {Comput. Phys. Commun.}\ }\textbf {\bibinfo
  {volume} {140}},\ \bibinfo {pages} {418} (\bibinfo {year} {2001})},\ \Eprint
  {https://arxiv.org/abs/hep-ph/0012260} {arXiv:hep-ph/0012260} \BibitemShut
  {NoStop}%
\bibitem [{\citenamefont {Mertig}\ \emph {et~al.}(1991)\citenamefont {Mertig},
  \citenamefont {B{\"o}hm},\ and\ \citenamefont {Denner}}]{Mertig:1990an}%
  \BibitemOpen
  \bibfield  {author} {\bibinfo {author} {\bibfnamefont {R.}~\bibnamefont
  {Mertig}}, \bibinfo {author} {\bibfnamefont {M.}~\bibnamefont {B{\"o}hm}},\
  and\ \bibinfo {author} {\bibfnamefont {A.}~\bibnamefont {Denner}},\ }\href
  {https://doi.org/10.1016/0010-4655(91)90130-D} {\bibfield  {journal}
  {\bibinfo  {journal} {Comput. Phys. Commun.}\ }\textbf {\bibinfo {volume}
  {64}},\ \bibinfo {pages} {345} (\bibinfo {year} {1991})}\BibitemShut
  {NoStop}%
\bibitem [{\citenamefont {Shtabovenko}\ \emph {et~al.}(2020)\citenamefont
  {Shtabovenko}, \citenamefont {Mertig},\ and\ \citenamefont
  {Orellana}}]{Shtabovenko:2020gxv}%
  \BibitemOpen
  \bibfield  {author} {\bibinfo {author} {\bibfnamefont {V.}~\bibnamefont
  {Shtabovenko}}, \bibinfo {author} {\bibfnamefont {R.}~\bibnamefont
  {Mertig}},\ and\ \bibinfo {author} {\bibfnamefont {F.}~\bibnamefont
  {Orellana}},\ }\href {https://doi.org/10.1016/j.cpc.2020.107478} {\bibfield
  {journal} {\bibinfo  {journal} {Comput. Phys. Commun.}\ }\textbf {\bibinfo
  {volume} {256}},\ \bibinfo {pages} {107478} (\bibinfo {year} {2020})},\
  \Eprint {https://arxiv.org/abs/2001.04407} {arXiv:2001.04407 [hep-ph]}
  \BibitemShut {NoStop}%
\bibitem [{\citenamefont {Maierh{\"o}fer}\ \emph {et~al.}(2018)\citenamefont
  {Maierh{\"o}fer}, \citenamefont {Usovitsch},\ and\ \citenamefont
  {Uwer}}]{Maierhofer:2017gsa}%
  \BibitemOpen
  \bibfield  {author} {\bibinfo {author} {\bibfnamefont {P.}~\bibnamefont
  {Maierh{\"o}fer}}, \bibinfo {author} {\bibfnamefont {J.}~\bibnamefont
  {Usovitsch}},\ and\ \bibinfo {author} {\bibfnamefont {P.}~\bibnamefont
  {Uwer}},\ }\href {https://doi.org/10.1016/j.cpc.2018.04.012} {\bibfield
  {journal} {\bibinfo  {journal} {Comput. Phys. Commun.}\ }\textbf {\bibinfo
  {volume} {230}},\ \bibinfo {pages} {99} (\bibinfo {year} {2018})},\ \Eprint
  {https://arxiv.org/abs/1705.05610} {arXiv:1705.05610 [hep-ph]} \BibitemShut
  {NoStop}%
\bibitem [{\citenamefont {Klappert}\ \emph {et~al.}(2021)\citenamefont
  {Klappert}, \citenamefont {Lange}, \citenamefont {Maierh{\"o}fer},\ and\
  \citenamefont {Usovitsch}}]{Klappert:2020nbg}%
  \BibitemOpen
  \bibfield  {author} {\bibinfo {author} {\bibfnamefont {J.}~\bibnamefont
  {Klappert}}, \bibinfo {author} {\bibfnamefont {F.}~\bibnamefont {Lange}},
  \bibinfo {author} {\bibfnamefont {P.}~\bibnamefont {Maierh{\"o}fer}},\ and\
  \bibinfo {author} {\bibfnamefont {J.}~\bibnamefont {Usovitsch}},\ }\href
  {https://doi.org/10.1016/j.cpc.2021.108024} {\bibfield  {journal} {\bibinfo
  {journal} {Comput. Phys. Commun.}\ }\textbf {\bibinfo {volume} {266}},\
  \bibinfo {pages} {108024} (\bibinfo {year} {2021})},\ \Eprint
  {https://arxiv.org/abs/2008.06494} {arXiv:2008.06494 [hep-ph]} \BibitemShut
  {NoStop}%
\bibitem [{\citenamefont {Laporta}(2000)}]{Laporta:2000dsw}%
  \BibitemOpen
  \bibfield  {author} {\bibinfo {author} {\bibfnamefont {S.}~\bibnamefont
  {Laporta}},\ }\href {https://doi.org/10.1142/S0217751X00002159} {\bibfield
  {journal} {\bibinfo  {journal} {Int. J. Mod. Phys. A}\ }\textbf {\bibinfo
  {volume} {15}},\ \bibinfo {pages} {5087} (\bibinfo {year} {2000})},\ \Eprint
  {https://arxiv.org/abs/hep-ph/0102033} {arXiv:hep-ph/0102033} \BibitemShut
  {NoStop}%
\bibitem [{\citenamefont {Liu}\ \emph {et~al.}(2018)\citenamefont {Liu},
  \citenamefont {Ma},\ and\ \citenamefont {Wang}}]{Liu:2017jxz}%
  \BibitemOpen
  \bibfield  {author} {\bibinfo {author} {\bibfnamefont {X.}~\bibnamefont
  {Liu}}, \bibinfo {author} {\bibfnamefont {Y.-Q.}\ \bibnamefont {Ma}},\ and\
  \bibinfo {author} {\bibfnamefont {C.-Y.}\ \bibnamefont {Wang}},\ }\href
  {https://doi.org/10.1016/j.physletb.2018.02.026} {\bibfield  {journal}
  {\bibinfo  {journal} {Phys. Lett. B}\ }\textbf {\bibinfo {volume} {779}},\
  \bibinfo {pages} {353} (\bibinfo {year} {2018})},\ \Eprint
  {https://arxiv.org/abs/1711.09572} {arXiv:1711.09572 [hep-ph]} \BibitemShut
  {NoStop}%
\bibitem [{\citenamefont {Liu}\ and\ \citenamefont {Ma}(2023)}]{Liu:2022chg}%
  \BibitemOpen
  \bibfield  {author} {\bibinfo {author} {\bibfnamefont {X.}~\bibnamefont
  {Liu}}\ and\ \bibinfo {author} {\bibfnamefont {Y.-Q.}\ \bibnamefont {Ma}},\
  }\href {https://doi.org/10.1016/j.cpc.2022.108565} {\bibfield  {journal}
  {\bibinfo  {journal} {Comput. Phys. Commun.}\ }\textbf {\bibinfo {volume}
  {283}},\ \bibinfo {pages} {108565} (\bibinfo {year} {2023})},\ \Eprint
  {https://arxiv.org/abs/2201.11669} {arXiv:2201.11669 [hep-ph]} \BibitemShut
  {NoStop}%
\bibitem [{\citenamefont {Lee}(2012)}]{Lee:2012cn}%
  \BibitemOpen
  \bibfield  {author} {\bibinfo {author} {\bibfnamefont {R.~N.}\ \bibnamefont
  {Lee}},\ }\href@noop {} {\  (\bibinfo {year} {2012})},\ \Eprint
  {https://arxiv.org/abs/1212.2685} {arXiv:1212.2685 [hep-ph]} \BibitemShut
  {NoStop}%
\bibitem [{\citenamefont {Lee}(2014)}]{Lee:2013mka}%
  \BibitemOpen
  \bibfield  {author} {\bibinfo {author} {\bibfnamefont {R.~N.}\ \bibnamefont
  {Lee}},\ }\href {https://doi.org/10.1088/1742-6596/523/1/012059} {\bibfield
  {journal} {\bibinfo  {journal} {J. Phys. Conf. Ser.}\ }\textbf {\bibinfo
  {volume} {523}},\ \bibinfo {pages} {012059} (\bibinfo {year} {2014})},\
  \Eprint {https://arxiv.org/abs/1310.1145} {arXiv:1310.1145 [hep-ph]}
  \BibitemShut {NoStop}%
\bibitem [{\citenamefont {Ma{\^i}tre}(2006)}]{Maitre:2005uu}%
  \BibitemOpen
  \bibfield  {author} {\bibinfo {author} {\bibfnamefont {D.}~\bibnamefont
  {Ma{\^i}tre}},\ }\href {https://doi.org/10.1016/j.cpc.2005.10.008} {\bibfield
   {journal} {\bibinfo  {journal} {Comput. Phys. Commun.}\ }\textbf {\bibinfo
  {volume} {174}},\ \bibinfo {pages} {222} (\bibinfo {year} {2006})},\ \Eprint
  {https://arxiv.org/abs/hep-ph/0507152} {arXiv:hep-ph/0507152} \BibitemShut
  {NoStop}%
\bibitem [{\citenamefont {Ma{\^i}tre}(2012)}]{Maitre:2007kp}%
  \BibitemOpen
  \bibfield  {author} {\bibinfo {author} {\bibfnamefont {D.}~\bibnamefont
  {Ma{\^i}tre}},\ }\href {https://doi.org/10.1016/j.cpc.2011.11.015} {\bibfield
   {journal} {\bibinfo  {journal} {Comput. Phys. Commun.}\ }\textbf {\bibinfo
  {volume} {183}},\ \bibinfo {pages} {846} (\bibinfo {year} {2012})},\ \Eprint
  {https://arxiv.org/abs/hep-ph/0703052} {arXiv:hep-ph/0703052} \BibitemShut
  {NoStop}%
\bibitem [{\citenamefont {Duhr}\ and\ \citenamefont
  {Dulat}(2019)}]{Duhr:2019tlz}%
  \BibitemOpen
  \bibfield  {author} {\bibinfo {author} {\bibfnamefont {C.}~\bibnamefont
  {Duhr}}\ and\ \bibinfo {author} {\bibfnamefont {F.}~\bibnamefont {Dulat}},\
  }\href {https://doi.org/10.1007/JHEP08(2019)135} {\bibfield  {journal}
  {\bibinfo  {journal} {JHEP}\ }\textbf {\bibinfo {volume} {08}},\ \bibinfo
  {pages} {135}},\ \Eprint {https://arxiv.org/abs/1904.07279} {arXiv:1904.07279
  [hep-th]} \BibitemShut {NoStop}%
\bibitem [{\citenamefont {Bauer}\ \emph {et~al.}(2002)\citenamefont {Bauer},
  \citenamefont {Frink},\ and\ \citenamefont {Kreckel}}]{Bauer:2000cp}%
  \BibitemOpen
  \bibfield  {author} {\bibinfo {author} {\bibfnamefont {C.}~\bibnamefont
  {Bauer}}, \bibinfo {author} {\bibfnamefont {A.}~\bibnamefont {Frink}},\ and\
  \bibinfo {author} {\bibfnamefont {R.}~\bibnamefont {Kreckel}},\ }\href
  {https://doi.org/10.1006/jsco.2001.0494} {\bibfield  {journal} {\bibinfo
  {journal} {J. Symb. Comput.}\ }\textbf {\bibinfo {volume} {33}},\ \bibinfo
  {pages} {1} (\bibinfo {year} {2002})},\ \Eprint
  {https://arxiv.org/abs/cs/0004015} {arXiv:cs/0004015} \BibitemShut {NoStop}%
\bibitem [{\citenamefont {Vollinga}\ and\ \citenamefont
  {Weinzierl}(2005)}]{Vollinga:2004sn}%
  \BibitemOpen
  \bibfield  {author} {\bibinfo {author} {\bibfnamefont {J.}~\bibnamefont
  {Vollinga}}\ and\ \bibinfo {author} {\bibfnamefont {S.}~\bibnamefont
  {Weinzierl}},\ }\href {https://doi.org/10.1016/j.cpc.2004.12.009} {\bibfield
  {journal} {\bibinfo  {journal} {Comput. Phys. Commun.}\ }\textbf {\bibinfo
  {volume} {167}},\ \bibinfo {pages} {177} (\bibinfo {year} {2005})},\ \Eprint
  {https://arxiv.org/abs/hep-ph/0410259} {arXiv:hep-ph/0410259} \BibitemShut
  {NoStop}%
\bibitem [{\citenamefont {Workman}\ \emph {et~al.}(2022)\citenamefont {Workman}
  \emph {et~al.}}]{ParticleDataGroup:2022pth}%
  \BibitemOpen
  \bibfield  {author} {\bibinfo {author} {\bibfnamefont {R.~L.}\ \bibnamefont
  {Workman}} \emph {et~al.} (\bibinfo {collaboration} {Particle Data Group}),\
  }\href {https://doi.org/10.1093/ptep/ptac097} {\bibfield  {journal} {\bibinfo
   {journal} {PTEP}\ }\textbf {\bibinfo {volume} {2022}},\ \bibinfo {pages}
  {083C01} (\bibinfo {year} {2022})}\BibitemShut {NoStop}%
\bibitem [{\citenamefont {Chen}\ \emph {et~al.}(2014)\citenamefont {Chen},
  \citenamefont {Ma}, \citenamefont {Zhang}, \citenamefont {Zhang},
  \citenamefont {Chen},\ and\ \citenamefont {Guo}}]{Chen:2014iwb}%
  \BibitemOpen
  \bibfield  {author} {\bibinfo {author} {\bibfnamefont {C.}~\bibnamefont
  {Chen}}, \bibinfo {author} {\bibfnamefont {W.-G.}\ \bibnamefont {Ma}},
  \bibinfo {author} {\bibfnamefont {R.-Y.}\ \bibnamefont {Zhang}}, \bibinfo
  {author} {\bibfnamefont {Y.}~\bibnamefont {Zhang}}, \bibinfo {author}
  {\bibfnamefont {L.-W.}\ \bibnamefont {Chen}},\ and\ \bibinfo {author}
  {\bibfnamefont {L.}~\bibnamefont {Guo}},\ }\href
  {https://doi.org/10.1140/epjc/s10052-014-3166-y} {\bibfield  {journal}
  {\bibinfo  {journal} {Eur. Phys. J. C}\ }\textbf {\bibinfo {volume} {74}},\
  \bibinfo {pages} {3166} (\bibinfo {year} {2014})},\ \Eprint
  {https://arxiv.org/abs/1409.4900} {arXiv:1409.4900 [hep-ph]} \BibitemShut
  {NoStop}%
\bibitem [{\citenamefont {Zhang}\ \emph {et~al.}(2016)\citenamefont {Zhang},
  \citenamefont {Duan}, \citenamefont {Ma}, \citenamefont {Zhang},\ and\
  \citenamefont {Chen}}]{Zhang:2015prr}%
  \BibitemOpen
  \bibfield  {author} {\bibinfo {author} {\bibfnamefont {Y.}~\bibnamefont
  {Zhang}}, \bibinfo {author} {\bibfnamefont {P.-F.}\ \bibnamefont {Duan}},
  \bibinfo {author} {\bibfnamefont {W.-G.}\ \bibnamefont {Ma}}, \bibinfo
  {author} {\bibfnamefont {R.-Y.}\ \bibnamefont {Zhang}},\ and\ \bibinfo
  {author} {\bibfnamefont {C.}~\bibnamefont {Chen}},\ }\href
  {https://doi.org/10.1140/epjc/s10052-016-3919-x} {\bibfield  {journal}
  {\bibinfo  {journal} {Eur. Phys. J. C}\ }\textbf {\bibinfo {volume} {76}},\
  \bibinfo {pages} {76} (\bibinfo {year} {2016})},\ \Eprint
  {https://arxiv.org/abs/1512.01879} {arXiv:1512.01879 [hep-ph]} \BibitemShut
  {NoStop}%
\end{thebibliography}%

\end{document}